\documentclass[conference]{IEEEtran}

\usepackage{graphicx}
\usepackage{lipsum}
\usepackage{hyperref}
\usepackage{cite}
\usepackage{pifont}
\usepackage{enumitem}
\usepackage{caption}
\usepackage{adjustbox}
\usepackage{subcaption}

\IEEEoverridecommandlockouts
\usepackage{atbegshi}
\AtBeginDocument{\AtBeginShipoutNext{\AtBeginShipoutDiscard}}

\begin{document}

\title{\LARGE Optimal Hybrid Transmit Beamforming for mm-Wave Integrated Sensing and Communication \vspace{-5mm}}
\author{\small
Jitendra~Singh, \textit{Graduate Student Member,~IEEE,}
Banda Naveen,
Suraj~Srivastava, \textit{Member,~IEEE,}
Aditya~K.~Jagannatham, \\ \textit{Senior Member,~IEEE}
and Lajos~Hanzo, \textit{Life Fellow,~IEEE}\vspace{-10mm}
}
\thanks{J. Singh B. Naveen, and A. K. Jagannatham are with the Department of Electrical Engineering, Indian Institute of Technology Kanpur, Kanpur, UP 208016, India (e-mail: jitend@iitk.ac.in; naveenb22@iitk.ac.in; adityaj@iitk.ac.in).
}
\thanks{S. Srivastava is with the Department of Electrical Engineering, Indian Institute of Technology Jodhpur, Jodhpur, Rajasthan 342030, India (email: surajsri@iitj.ac.in).}
\thanks{L. Hanzo is with the School of Electronics and Computer Science, University of Southampton, Southampton SO17 1BJ, U.K. (e-mail: lh@ecs.soton.ac.uk).
}

\maketitle
\begin{abstract}
A hybrid beamformer (HBF) is designed for integrated sensing and communication (ISAC)-aided millimeter wave (mmWave) systems. The ISAC base station (BS), relying on a limited number of radio frequency (RF) chains, supports multiple communication users (CUs) and simultaneously detects the radar target (RT). To maximize the probability of detection (PD) of the RT, and achieve rate fairness among the CUs, we formulate two problems for the optimization of the RF and baseband (BB) transmit precoders (TPCs): PD-maximization (PD-max) and geometric mean rate-maximization (GMR-max), while ensuring the quality of services (QoS) of the RT and CUs. Both problems are highly non-convex due to the intractable expressions of the PD and GMR and also due to the non-convex unity magnitude constraints imposed on each element of the RF TPC. To solve these problems, we first transform the intractable expressions into their tractable counterparts and propose a power-efficient bisection search and majorization and minimization-based alternating algorithms for the PD-max and GMR-max problems, respectively. Furthermore, both algorithms optimize the BB TPC and RF TPCs in an alternating fashion via the successive convex approximation (SCA) and penalty-based Riemannian conjugate gradient (PRCG) techniques, respectively. Specifically, in the PRCG method, we initially add all the constraints except for the unity magnitude constraint to the objective function as a penalty term and subsequently employ the RCG method for optimizing the RF TPC.
Finally, we present our simulation results and compare them to the benchmarks for demonstrating the efficacy of the proposed algorithms.

\end{abstract}
\maketitle
\section{\uppercase{INTRODUCTION}}
\IEEEPARstart{I}{ntegrated} sensing, and communication (ISAC) systems are at the forefront \cite{ISAC_MIMO_6,ne_1} of the recent era of wireless technology. The integration of sensing and communication functionalities in such systems offers several practical advantages, which can drive a wide range of applications and unlock cutting-edge capabilities in next-generation (NG) wireless communication \cite{ji_3}. 
Briefly, ISAC systems improve hardware efficiency by enabling the use of a single platform for sensing the radar targets (RTs) as well as for communication. By doing so, one can reduce the size, cost, and power consumption of the hardware in ISAC systems in comparison to conventional systems that require separate hardware for sensing and communication \cite{ji_4,ji_5,ji_6}.

Furthermore, due to the bandwidth crunch in the sub-6 GHz frequency band, the NG wireless networks might migrate to the millimeter wave (mmWave) band, which spans the frequency range $30$ to $300$ GHz \cite{ji_mm_1}. Moreover, both the hardware architecture and signal processing necessary for sensing the RTs are similar to those of mmWave systems. Therefore, the integration of ISAC and mmWave capabilities can reap the benefits of a wide frequency band and mutual signal processing advances, while requiring only moderate hardware changes in the existing mmWave systems. 

It is crucial to employ optimal beamforming techniques to realize the full potential of ISAC-aided mmWave systems for sensing and communication. Furthermore, it is important to note that the conventional fully digital beamforming (FDB) techniques of sub-6 GHz systems \cite{new_100,new_101,new_102} necessitate an individual RF chain (RFC) for each antenna, thus rendering them costly and power-inefficient. Hence one cannot employ such techniques in mmWave-aided ISAC systems.
To address this, numerous authors have investigated hybrid beamforming (HBF) designs, which require only a few RFCs and are therefore eminently suitable for mmWave-aided ISAC systems \cite{jite_1,mmISAC_1,mmISAC_2,HBF_8,mmISAC_3,mmISAC_4,mmISAC_5,mmISAC_6,mm_OMP}. In HBF, the signal processing of the transmit precoder (TPC) is divided into the digital baseband (BB) and analog RF TPCs, where analog RF processing comprises phase shifters designed for beam steering in the RF domain. Furthermore, several endeavors focused on jointly optimizing the beamforming designs of the ISAC systems for the maximization of either the sum rate or minimum rate (MR) of the users. However, the maximization of the sum rate results in rate unfairness, while the maximization of the MR results in a reduced sum rate. To this end, the recent geometric mean (GM) rate-based communication metric \cite{Without_ISAC_3} has demonstrated superior capability to achieve rate fairness without unduly compromising the sum rate of the system, thus delivering a compelling trade-off.
Before we delve into the details of our framework, in the next subsection, we discuss the rich prior literature in this area and the motivation behind our investigation of novel techniques designed for mmWave-aided ISAC systems.

\subsection{Literature review} \label{literature review}
The authors of \cite{SUB6_ISAC_1,SUB6_ISAC_2,SUB6_ISAC_3,SUB6_ISAC_4,SUB6_ISAC_5,new_100,new_101,new_102,new_105} 
designed pioneering beamforming techniques for ISAC systems operating in the sub-6 GHz frequency band. 
Specifically, the authors of \cite{new_100} have considered both separated and shared antenna deployments, optimizing TPCs to match radar beampatterns, while ensuring communication users (CUs) performance. 
Liu \textit{et al.} \cite{new_101} have explored symbol-level TPC in ISAC systems to enhance beamforming flexibility, ensuring improved instantaneous transmit beampatterns and better CUs performance.
Unlike conventional block-level precoding techniques, authors of \cite{new_102} proposed a symbol-level TPC approach to enhance the instantaneous transmit beampatterns for radar sensing while maintaining communication performance, demonstrating the potential of advanced beamforming techniques in ISAC systems.
Ding \textit{et al.} \cite{SUB6_ISAC_1} maximized the communication performance of the CUs by minimizing the multi-user interference (MUI), while guaranteeing the probability of detection (PD) of the RT detection in ISAC systems. By contrast, Ashraf and Tan \cite{SUB6_ISAC_2} maximized the sensing performance of the RT by maximizing the PD, while constraining the minimum signal-to-interference and noise ratio (SINR) requirements of the CUs. 
Furthermore, Liu \textit{et al.} \cite{SUB6_ISAC_3} considered the Cramér-Rao bound (CRB) as the performance metric for sensing the RT and minimized the CRB of the parameters in both point as well as extended RTs, while guaranteeing the minimum SINR requirements of the CUs. 
Bazzi and Chafii \cite{SUB6_ISAC_4} considered realistic imperfect channel state information (CSI) for the CUs and maximized the received SNR of the RT under probabilistic SINR outage constraints. 
On the other hand, Zargari \textit{et al.} in \cite{SUB6_ISAC_5} maximized the sum rate of the CUS, while meeting a minimum beampattern gain for the RTs. Briefly, they optimized the beamformer by exploiting the Lagrangian and Riemannian manifold optimization principles.
As a further advance, artificial intelligent (AI)-driven techniques for ISAC systems, such as deep unfolding learning and AI-based iterative optimization \cite{new_106}, have been employed to optimize RIS-aided ISAC by minimizing interference, enhancing waveform design efficiency, and solving complex non-convex problems in secure and robust transmission \cite{new_107}.

However, the above studies employ the FDB design, which poses a significant drawback in the context of mmWave-aided ISAC systems due to the requirement for a large number of RFCs. 
In order to circumvent this issue, the authors of \cite{mmISAC_1,mmISAC_2,mmISAC_3,mmISAC_4,mmISAC_5,mmISAC_6,mmWave_ISAC_1,mmWave_ISAC_3,mmWave_ISAC_2} have proposed diverse HBF designs for mmWave ISAC systems, which aimed for significantly reducing the number of RFCs. 
Qi \textit{et al.} \cite{mmISAC_1} proposed a two-stage HBF, which minimizes the error between the ideal and transmit beam pattern gain for the radar RT, while considering the minimum SINR requirement of the CU as a constraint. Furthermore, Barneto \textit{et al.} \cite{mmISAC_2} consider the full duplex paradigm in a mmWave-aided ISAC system, wherein they optimize the components of the HBF BB and RF TPCs for maximizing the beamforming power towards the RT, while constraining the beamforming power toward the CUs. As a further advance, Yu \textit{et al.} \cite{mmISAC_5} designed HBF for the mmWave ISAC-assisted internet of vehicles (IoVs), wherein they formulate the joint HBF design problem as the weighted summation of the communication beamforming error and radar beamforming error. To solve the problem, the authors therein proposed a pair of novel methods: fast Riemannian manifold optimization (FRMO) and adaptive particle swarm optimization (APSO). Moreover, the authors of \cite{mmWave_ISAC_1,mmWave_ISAC_3} optimized the HBF of mmWave-aided ISAC systems, which leads to the maximization of the weighted sum rate of the CUs. 
Specifically, Gong \textit{et al.} \cite{mmWave_ISAC_1} evaluated the CRB for the estimation of the DoA, while Zhou \textit{et al.} \cite{mmWave_ISAC_3} employed the beampattern gain towards the RTs as the constraint for their radar performance optimization. The authors of \cite{mmWave_ISAC_1} therein equivalently transformed the non-convex problem into a convex one via the weighted
minimum mean square error (WMMSE) method and subsequently employed the alternating optimization technique for optimizing the BB and RF TPCs. 
Furthermore, Wang et al. \cite{mmWave_ISAC_2} investigated the partially-connected hybrid MIMO architecture for mmWave ISAC systems, where they aimed to minimize the CRB for angle of departure (AoD) estimation while ensuring the SINR constraints of the CUs. Unlike the fully-connected architecture, the partially-connected structure employs a block-diagonal RF TPC matrix, which reduces hardware complexity at the cost of some beamforming flexibility.

\begin{table*}[t!]
    \centering
    \caption{Summary of literature survey on mmWave MIMO ISAC systems} \label{tab:lit_rev}
    \begin{adjustbox}{width=\linewidth}
\begin{tabular}{|l|c|c|c|c|c|c|c|c|c|c|c|c|c|c|c|c|c|c|c|c|c|c|c|c|c|}
    \hline
   &\cite{new_100} &\cite{new_101} &\cite{new_102}  &\cite{mmISAC_1} &\cite{mmISAC_2}  &\cite{mmISAC_5} &\cite{Without_ISAC_3}  &\cite{SUB6_ISAC_1}  &\cite{SUB6_ISAC_2}   &\cite{SUB6_ISAC_4}  &\cite{SUB6_ISAC_5}  &\cite{mmWave_ISAC_1}  &\cite{mmWave_ISAC_3}  & \cite{mmWave_ISAC_2} &\cite{ji_2}   &\cite{ji_PD_3}  &\cite{Without_ISAC_2}  & Proposed\\ [0.5ex]
 \hline
mmWave ISAC  &  & &  &\checkmark    &\checkmark     &\checkmark   &   &   &   &  &   &   &  &\checkmark  &      &   &  &\checkmark \\ \hline
 
HBF &  &  &  &\checkmark    &\checkmark  &\checkmark   &   &   &  &   &   &\checkmark   &\checkmark  &\checkmark  &  &    &\checkmark   &\checkmark  \\ \hline

PD-max &  &  &  & &  &    &    &   &\checkmark  &\checkmark  &  & & &  &  &\checkmark       &  &\checkmark  \\ \hline

GMR-max &  &  & & &   &   &\checkmark  &   &   &  &  &   &  & & &      &\checkmark  &\checkmark   \\ \hline

SINR as QoS constraint &  &\checkmark &\checkmark  &\checkmark   &    &  &  &  &\checkmark    &\checkmark  &\checkmark  & &  &\checkmark & &\checkmark   &    &\checkmark  \\ \hline

PD as QoS constraint & & & &  &    &    &  &\checkmark  &   &  &   &  & &  &\checkmark   &     &    &\checkmark  \\ \hline

RCG  &\checkmark  &  &\checkmark &  &  &\checkmark  &   &  &  &  &\checkmark  &\checkmark  & &  &  &   &      &\checkmark  \\ \hline
 
SCA  &  & & &   &  & &  &  &  &  &  &\checkmark  &  &\checkmark  &  &      &   &\checkmark   \\ \hline

MM & &\checkmark &\checkmark & &  &   &  &  &    &   &   &   &\checkmark  &  &  &      &\checkmark  &\checkmark  \\ \hline

{\bf PD-max in mmWave ISAC} &  & & &  &    &   &  &   &   &  &   &   &   &  &     &   &  &\checkmark   \\ \hline

{\bf GMR-max in mmWave ISAC} & & & &  &   &  &   &  &   &  &   &  &  & &     &   &   &\checkmark  \\ \hline

{\bf Bi-Alt} & & & & &   &  &  &  & &  & &  & & &  &   &  &\checkmark \\ \hline
 
{\bf MM-Alt} &  & &\checkmark &  &  &   &   &   &  &   &  &  &  & &    &   &  &\checkmark \\ \hline

{\bf Penalty-based RCG} & & &\checkmark  &  & &  &  &  &   &   &   &    &   &   &  & &   &\checkmark \\ \hline

\end{tabular}
\end{adjustbox}
\end{table*}

However, the key metric for the sensing performance is the PD, while is still unexplored in the context of mmWave-enabled ISAC systems. By contrast, the authors of \cite{ji_2,SUB6_ISAC_1,ji_PD_2,ji_PD_3} investigated the PD in the context of sub-6 GHz systems and obtained the expression of the PD via the generalized likelihood ratio test (GLRT). 
Furthermore, the existing literature on mmWave ISAC systems has not as yet explored the principle of rate fairness for the CUs, which is also a key performance metric. To elaborate briefly, maximizing the sum rate typically assigns the most resources to the CUs having the best channel, while assigning a near-zero rate to the CUs having low channel quality, especially as the radar performance improves. However, in order to ensure rate fairness, significant research efforts have been dedicated to the beamforming design beyond ISAC, focusing either on maximizing the minimum CU rate or the GM rate of the CU \cite{Without_ISAC_2, Without_ISAC_3}. 
The authors of \cite{Without_ISAC_2} proposed a cutting-edge HBF design by solving a max-min rate (MMR) optimization problem in a mmWave system. Moreover, Yu \textit{et al.} \cite{Without_ISAC_3} considered a RIS-aided wireless system and jointly optimized the active and passive beamformers by maximizing the GM rate of the CUs. Furthermore, the novel transformations of the objectives in \cite{Without_ISAC_2, Without_ISAC_3} were achieved using the theory of majorization-minimization (MM), and subsequently, closed-form expressions are derived for the optimal solutions, which renders these studies potent in practical deployments.
Compared to MMR optimization, which prioritizes the worst-case user, maximization of GM rates balances fairness and overall performance. In mmWave ISAC systems with directional beams and blockages, it prevents resource domination by a single user while improving weaker users' rates. This enhances spectral efficiency and mitigates performance degradation seen in strict MMR formulations, making it a practical choice for ISAC deployments.

Motivated by these facts, we investigate the PD- and GM rate-maximization problem in an mmWave ISAC system. To the best of our knowledge, this is the first paper that investigates HBF for the optimization of the BB and RF TPCs and maximizes the PD as well as GM rate obtained for the sensing and communication subsystems, respectively. To this end, we commence with the system model of a downlink mmWave-aided ISAC system, where an ISAC BS provides communication services to CUs while simultaneously detecting the RT based on the received echo signal. Our novel contributions are boldly and explicitly contrasted to the existing literature in Table  \ref{tab:lit_rev} and are further described next.

\subsection{Contributions of this work}\label{contributions}
\begin{itemize}
    \item We develop a rigorous framework for jointly evaluating sensing and communication performance by employing the generalized likelihood ratio test (GLRT) to determine the PD of the RT and deriving a tractable rate expression for the CUs. These formulations serve as the foundation for the proposed joint optimization framework, enabling efficient resource allocation and beamforming design while ensuring a balanced trade-off between sensing and communication objectives.

    \item Based on the system model considered as well as on the sensing and communication performance metrics, we formulate a pair of optimization problems, namely PD-max and GMR-max. For the PD-max problem, we aim for maximizing the PD of the RT, while meeting the minimum SINR requirements of the CUs. By contrast, in the GMR-max problem, we focus on maximizing the GM rate of the CUs, while satisfying the minimum PD of the RT. 
    Both problems are highly non-convex due to the intractable PD and GM rate expressions, owing to the non-convex unit modulus constraint, and to the tightly coupled optimization variables. 
    \item To solve the first problem, we propose a bisection search-based alternating (Bi-Alt) algorithm, where we first transform the original problem to an equivalent power minimization problem by introducing a slack variable for the PD. Subsequently, we optimize the BB and RF TPCs in an alternating fashion and update the slack variable via a bisection search. Specifically, we adopt the SCA method for optimizing the BB TPC, whereas we propose a novel penalty-based Riemannian conjugate gradient (PRCG) algorithm for optimizing the RF TPC.
    \item For the second GMR-max problem, we first transform the intractable GMR-max problem into a tractable weighted sum rate maximization problem, which is still non-convex due to the non-convex rate expression. To address this challenge, we propose a majorization and minimization-based alternating (MM-Alt) algorithm, where we obtain the convex surrogate function for the non-convex rate function via the MM technique and optimize both BB and RF TPCs in an alternating fashion via the SCA and PRCG methods, respectively.
    
    
   
    \item The performance of the proposed HBF designs is characterized via simulations and also compared to the pertinent benchmarks, which validates the efficiency of the proposed methods.
\end{itemize}

\subsection{Notations}\label{notation}
We use the following notations throughout the
paper: $\mathbf{A}$,  $\mathbf{a}$, and $a$ represent a matrix, a vector, and a scalar quantity, respectively.
The $(i,j)$th element, and Hermitian of matrix $\mathbf{A}$ are denoted by  $\mathbf{A}{(i,j)}$, and $\mathbf{A}^H$, respectively. The trace, Frobenius
norm and vectorization of a matrix $\mathbf{A}$ are represented as $\mathrm{tr}\left(\cdot\right)$, $\left\vert\left\vert \mathbf{A} \right\vert\right\vert_F$ and $\text{vec}\left (\cdot\right)$. The expectation operator is represented as $\mathbf{E}\{\cdot\}$; the real part of a quantity is denoted by ${\Re\left (\cdot\right)}$; ${\mathbf I}_M$ denotes an $M \times M$ identity matrix; the symmetric complex Gaussian distribution of the mean $\mathbf{\mu}$ and the covariance matrix $\mathbf{\sigma^2}$  is represented as ${\cal CN}(\mathbf{\mu}, \mathbf{\sigma^2})$.The operators $\odot$ and $\otimes$ denote the Hadamard product and  Kronecker product, respectively.
\section{\uppercase{System Model}}\label{System Model}
\begin{figure}[t]
        \vspace{-1.3cm}
         \centering 
                \includegraphics[width=\linewidth]{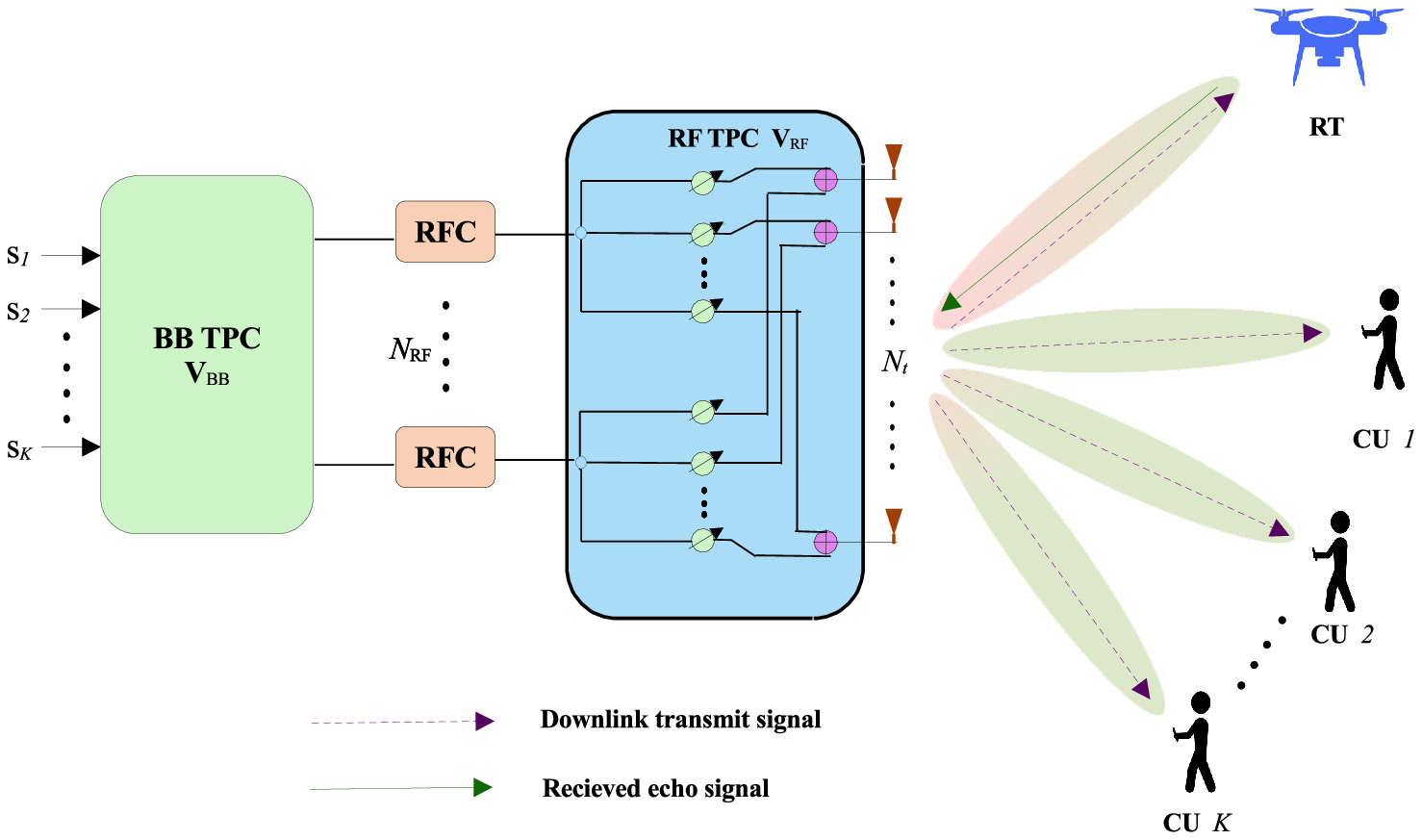}
                \vspace{-1.5cm}
                 \caption{Illustration of a mmWave-aided ISAC system.}
                \label{fig:sys_1}
                \vspace{-5mm}
                \end{figure}
        
We consider the mmWave ISAC downlink as illustrated in Fig. \ref{fig:sys_1}, where an ISAC base station (BS) serves $K$ CUs and simultaneously senses a point-like RT. The ISAC BS is equipped with $N_\mathrm{t}$ transmit antennas, whereas each CU has a single antenna. To reduce both the cost and hardware power consumption, the ISAC BS exploits a fully-connected hybrid MIMO architecture, in which the overall TPC is split into digital BB and analog RF domain TPCs. Both the TPCs are connected via a limited number of RFCs, $N_\mathrm{RF}$, and the RF TPC is connected to the antenna elements via phase shifters. To ensure the feasibility of the beamforming design problem, we assume the condition $K < N_\mathrm{RF} << N_\mathrm{t}$.

\subsection{Signal model}
Let us define the data stream vector as $\mathbf{s}=[s_1, \ldots, s_K]\in {\mathbb C}^{{K} \times 1}$, where $s_k$ denotes the data for the $k$th CU. 
We assume that the data symbols are statistically independent with zero means, i.e., $\mathbb{E}\{\mathbf{s}\mathbf{s}^H\}=\mathbf{I}_K$ and $\mathbb{E}\{\mathbf{s}_k\}=0, \forall k$.
Following the HBF design, the signal vector $\mathbf{s}$ is first processed by the BB TPC $\mathbf{V}_\mathrm{BB}=[\mathbf{v}_{{\rm BB},1}, \hdots,\mathbf{v}_{{\rm BB}, K}]\in {\mathbb C}^{{N_\mathrm{RF}} \times K}$ followed by the RF TPC $\mathbf{V}_\mathrm{RF}\in {\mathbb C}^{{N_\mathrm{t}} \times {N_\mathrm{RF}}}$, where $\mathbf{v}_{{\rm BB},k}$ represents the BB TPC vector corresponding to the $k$th CU. Consequently, the downlink signal $\mathbf {x} \in \mathbb{C}^{N_\mathrm{t}\times 1}$ transmitted from the ISAC BS is given by
\begin{equation} \label{eqn:tx_signal}
\mathbf{x}=\mathbf{V}_{\rm RF}\mathbf{V}_{\rm BB}\mathbf{s} = \mathbf{V}_{\rm RF}\sum_{k=1}^K\mathbf{v}_{{\rm BB},k}{s}_{k}.
\end{equation}
Embracing the principle of ISAC technology, the transmitted signal (\ref{eqn:tx_signal}) serves a dual role in downlink communication as well as radar detection. Therefore, we aim for efficiently designing $\mathbf{V}_{\rm BB}$ and $\mathbf{V}_{\rm RF}$ for a favorable performance tradeoff for RT detection and downlink communication. To this end, we define the radar and communication performance metrics in the subsequent subsections.
\subsection{Metric for radar performance}
We assume that the RT is located at a known distance and direction, denoted by $(r, \theta)$. Furthermore, a colocated MIMO radar is employed at the ISAC BS for detection, where the same antenna array elements are used for transmitting and receiving the radar signals.
When considering a uniform linear array (ULA) at the ISAC BS having a spacing of half-wavelength between the nearby antenna elements, the steering vector $\mathbf{a}_\mathrm{BS}(\theta)\in \mathbb{C}^{N_\mathrm{t}\times 1}$ can be expressed as 
\begin{equation}
\mathbf{a}_\mathrm{BS}(\theta) = \frac{1}{\sqrt{N_\mathrm{t}}}\left[1, e^{j\pi \sin(\theta)}, \ldots, e^{j(N_{\mathrm{t}}-1)\pi \sin(\theta)}\right]^T. 
\end{equation}
Furthermore, upon considering a clutter-free environment, the echo signal $\mathbf{y}_\mathrm{r} \in \mathbb{C}^{N_\mathrm{t} \times 1}$ received at the ISAC BS from the RT is expressed as
\begin{equation} \label{eqn:rad_rec}
\mathbf{y}_\mathrm{r}= \alpha{\mathbf{A}}(\theta)\mathbf{V}_{\rm RF}\mathbf{V}_{\rm BB}\mathbf{s} + \mathbf{v}, 
\end{equation}
where ${\mathbf{A}}(\theta) = \mathbf{a}_\mathrm{BS}(\theta)\mathbf{a}^T_\mathrm{BS}(\theta)\in \mathbb{C}^{N_\mathrm{t}\times N_\mathrm{t}}$ represents the effective radar channel matrix, and $\alpha$ denotes the radar cross-section (RCS) of the RT. Moreover, $\mathbf{v} \in \mathbb{C}^{N_\mathrm{t} \times 1}$ is the complex additive white Gaussian noise (AWGN) with distribution $\mathbf{v} \sim \mathcal{CN}(\mathbf{0}, \sigma_{v}^2 \mathbf{I}_{N_{t}})$, where $\sigma_{v}^2$ is the variance of the noise signal. 
For the radar performance metric, we aim for maximizing the PD by radiating the transmitted signal power toward the RT. The transmit power radiated in the direction of $\theta$, is given by
\begin{equation} \label{eqn:radar_tp}
\mathbf{P}(\theta)
=\sum_{k=1}^{K} \left| \mathbf{a}^{H}_\mathrm{BS}({\theta})\mathbf{V}_{\rm RF}\mathbf{v}_{{\rm BB}, k} \right|^2.
\end{equation}

Furthermore, to evaluate the sensing performance, we formulate an equivalent binary hypothesis testing problem for the RT located at $\theta$ based on the received echo signal (\ref{eqn:rad_rec}) as follows
\begin{equation} \label{eqn:CBHT_2} 
 \begin{cases} \displaystyle \mathcal {H}_{0}: &\alpha{\mathbf{A}}(\theta)\mathbf{V}_{\rm RF}\mathbf{V}_{\rm BB}\mathbf{s} =\mathbf{0}, ~\left[\mathbf{y}_\mathrm{r} = \mathbf{v}\right]\\ 
\displaystyle \mathcal {H}_{1}: &\alpha{\mathbf{A}}(\theta)\mathbf{V}_{\rm RF}\mathbf{V}_{\rm BB}\mathbf{s} \neq \mathbf{0}, ~\left[\mathbf{y}_\mathrm{r} = \alpha{\mathbf{A}}(\theta)\mathbf{V}_{\rm RF}\mathbf{V}_{\rm BB}\mathbf{s} + \mathbf{v}\right],\end{cases}
\end{equation}
where $\mathcal {H}_0$ and $\mathcal {H}_1$ are the null and alternate hypotheses, which represent the presence and absence of the RT in the detection range of the radar, respectively. Consequently, we adopt the generalized likelihood ratio test (GLRT) for the above problem (\ref{eqn:CBHT_2}), in which the unknown parameters $(\alpha,\theta)$ are substituted by their maximum likelihood estimates $(\hat{\alpha}_{\rm ML},\hat{\theta}_{\rm ML})$. Thus, the corresponding distribution of the optimal detector is given by \cite{SUB6_ISAC_2}
\begin{equation} \label{eqn:OD} 
L(\hat{\alpha}_{\rm ML},\hat{\theta}_{\rm ML}) \sim \begin{cases} \displaystyle \mathcal {H}_{0}: \chi_{2}^{2},\\ \displaystyle \mathcal {H}_{1}: \chi_{2}^{\prime 2}(\rho), \end{cases}
\end{equation}
where $\chi_{2}^{2}$ and $\chi_{2}^{\prime 2}(\rho)$ represent the central and noncentral chi-squared distributions with two degrees of freedom, respectively. Furthermore, $\rho$ denotes the non-central parameter \cite{SUB6_ISAC_1}, which is given by
\begin{equation} \label{eqn:NCP} 
\rho = \mu\left\vert\mathbf{P}(\theta)\right\vert^2,
\end{equation}
where $\mu = \frac{\left\vert\alpha\right\vert^2}{\sigma_{v}^2}$. Note that the non-central parameter in (\ref{eqn:NCP}) inherently depends on the number of receive antennas through $\left\vert\mathbf{P}(\theta)\right\vert$ in equation (\ref{eqn:radar_tp}).
Subsequently, upon employing the Neyman-Pearson criterion, the probability of detection, $P_\mathrm{D}$, of the RT for a given probability of false alarm $P_\mathrm{FA}$ becomes
\begin{subequations} \label{eqn:PD} 
\begin{align}
P_\mathrm{D} &= \mathcal{Q}_{\chi_{2}^{2}(\rho)} \Big (\mathcal{F}_{\chi_{2}^{2}}^{-1}\big(1 - P_\mathrm{ {FA}}\big)\Big),\\
&=  1 - \mathcal{F}_{\chi _{2}^{2}(\rho)} \Big(\mathcal{F}_{\chi_{2}^{2}}^{-1}\big(1 - P_\mathrm{{FA}}\big) \Big),
\end{align}
\end{subequations}
where ${Q}_{\chi_{2}^{2}(\rho)}(\cdot)$ denotes the tail probability, and ${F}_{\chi_{2}^{2}(\rho)}(\cdot)$ is the cumulative distribution function (CDF) of the noncentral chi-square distribution. 
Hence, this paper utilizes the above probability of detection, given by (\ref{eqn:PD}), to evaluate the radar performance of the system.

\subsection{Metric for communication performance}
Let us denote the mmWave MISO channel vector between the ISAC BS and $k$th CU by $\mathbf{h}^H_k \triangleq [h^*_{1,k}, \ldots, h^*_{N_\mathrm{t},k}] \in \mathbb{C}^{1 \times N_\mathrm{t}}$. To model the mmWave channel $\mathbf{h}^H_k$, we adopt the popular Saleh-Valenzuela model \cite{jite_1,jite_2}, which is expressed as
\begin{equation} \label{eqn:channel}
\mathbf{h}^H_{k}= \sqrt{\frac{N_\mathrm{t}}{{L_{k}}}}\sum_{\ell=1}^{L_{k}}\beta_{\ell,k} \mathbf{a}^H_\mathrm{BS}(\theta_{\ell,k}), 
\end{equation}
where $L_k$ is the number of multi-path components for the $k$th CU, and $\beta_{\ell,k}$ represents the complex path gain for the $\ell$th path of the $k$th CU. Furthermore, the quantity $\theta_{\ell,k}$ represents the azimuth angle of departure (AoD). 
Upon assuming the full CSI\footnote{In practice, CSI is obtained via channel estimation, with users feeding back CSI in frequency division duplex (FDD) systems or the BS estimating it via channel reciprocity in time division duplex (TDD) systems. Estimation and quantization errors may exist, in which case robust beamforming designs (see, e.g., \cite{SUB6_ISAC_4}) can be applied.} at the ISAC BS and CUs, the signal $y_k$ received at the $k$th CU is given by
\begin{subequations}\label{eqn:com_rec}
\begin{align}
y_k =&\mathbf{h}_k^H\mathbf{V}_\mathrm{RF}\mathbf{V}_\mathrm{BB}\mathbf{s} + n_k, \\
=&\mathbf{h}_k^H\mathbf{V}_\mathrm{RF}\mathbf{v}_{\mathrm{BB}, k}s_k + \sum_{i=1, i \neq k}^K\hspace{-0.3cm}\mathbf{h}_k^H \mathbf{V}_\mathrm{RF}\mathbf{v}_{\mathrm{BB},i}s_i+ n_k,
\end{align}
\end{subequations}
where $n_k$ is the independent and identically distributed (i.i.d.) complex AWGN at the $k$th CU having the distribution of $n_k\sim\mathcal{CN}({0}, \sigma^2_k)$. 
Based on (\ref{eqn:com_rec}), the signal-to-interference-plus-noise ratio (SINR) of the $k$th CU is given by
\begin{equation}\label{eqn:SINR}
\begin{aligned}
\mathbf{\gamma}_k(\mathbf{V}_\mathrm{RF},\mathbf{V}_\mathrm{BB}) = &\frac{\left\vert\mathbf{h}_k^H \mathbf{V}_\mathrm{RF}\mathbf{v}_{\mathrm{BB},k}\right\vert^2}
{\sum\limits_{i\neq k}^{K}{\left\vert\mathbf{h}_k^H \mathbf{V}_\mathrm{RF}\mathbf{v}_{\mathrm{B B},i}\right\vert^2} + \sigma_{k}^{2}}.
\end{aligned}
\end{equation}
Thus, the maximum achievable transmission rate of the $k$th CU is expressed in bits/s/Hz as follows
\begin{equation}\label{eqn:shanon}
R_k\big(\mathbf{V}_\mathrm{RF},\mathbf{V}_\mathrm{BB}\big)=\log_2{\Big(1+\gamma_k\big(\mathbf{V}_\mathrm{ RF},\mathbf{V}_\mathrm{BB}\big)\Big)}, \forall k.
\end{equation}
Based on the radar and communication performance metrics as discussed in the above subsections, we further describe the problem formulation in the next subsection.
\subsection{Problem formulation}\label{problem formulation}
This paper aims for jointly optimizing the RF TPC $\mathbf{V}_\mathrm{RF}$ and BB TPC $\mathbf{V}_\mathrm{BB}$ for the mmWave ISAC system. 
For joint optimization, we consider the following pair of criteria: 1) PD maximization (PD-max); 2) GMR maximization (GMR-max), which correspond to the maximization of the RT performance and the rate-fairness among CUs, respectively. 
Specifically, for the PD-max optimization problem, we maximize the PD of the RT under limited available power, while guaranteeing the minimum SINR requirement of the CUs. The corresponding PD-max problem is formulated as
\begin{subequations}\label{PD_OP:1}
\begin{align} 
\mathcal{P}_1:\hspace{20mm}&\mathop {\max }\limits_{\mathbf{V}_\mathrm{RF}, \mathbf{V}_\mathrm{BB}} \quad  P_\mathrm{D} \label{PD_OF:1}\\
&\text {s.t.} \quad \gamma_k(\mathbf{V}_\mathrm{RF},\mathbf{V}_\mathrm{BB}) \geq \Gamma_k, \forall k, \label{cons:SINR}\\ 
&\qquad\|\mathbf{V}_\mathrm{RF}\mathbf{V}_\mathrm{BB}\|_F^2\leq P_\mathrm{t},\label{cons:TP}\\
&\qquad \left\vert\mathbf{V}_\mathrm{RF}(i,j)\right\vert = 1, \forall i, j, \label{cons:RF}
\end{align}
\end{subequations}
where $\Gamma_k$ and $P_\mathrm{t}$ are the minimum SINR requirement of the $k$th CU and the transmit power available at the ISAC BS, respectively. Furthermore, the constraint (\ref{cons:RF}) represents the unity magnitude constraint at each element of the RF TPC due to the phase shifters. 

To ensure rate fairness among the CUs, we maximize the geometric mean rate of the CUs in the second optimization problem, while satisfying a minimum PD requirement in terms of the desired sensing performance. 
Thus, by defining the GMR as $f_\mathrm{GM}(\mathbf{V}_\mathrm{RF},\mathbf{V}_\mathrm{BB}) \triangleq  \left(\prod_{k=1}^K R_k(\mathbf{V}_\mathrm{RF}, \mathbf{V}_\mathrm{BB})\right)^{\frac{1}{K}}$, the corresponding GMR-max problem is formulated as follows
\begin{subequations}\label{GMR_OP:1}
\begin{align} 
\mathcal{P}_2: \hspace{8mm} 
&\mathop {\max}\limits_{\mathbf{V}_\mathrm{RF}, \mathbf{V}_\mathrm{BB}} \quad  f_\mathrm{GM}(\mathbf{V}_\mathrm{RF},\mathbf{V}_\mathrm{BB}) \label{GMR_OF:1}\\
&\quad \mathrm{s.\hspace{1mm} t.} \quad  \quad P_\mathrm{D} \geq P_\mathrm{th},\label{cons:Pd}\\ 
&\qquad \quad \quad  \text{(\ref{cons:TP}), and (\ref{cons:RF})},
\end{align}
\end{subequations}
where $P_\mathrm{th}$ denotes the specified threshold imposed on the PD of the RT for a given false alarm probability $P_\mathrm{FA}$. 

To account for dynamic RT scenarios, both the formulated problems (\ref{PD_OP:1}) and (\ref{GMR_OP:1}) are also applicable for moving RT tracking by steering the sensing beam toward the estimated or predicted direction \cite{new_105}. While localization errors may affect beam alignment and PD, adaptive tracking techniques such as Kalman or particle filtering can mitigate these effects \cite{new_103,new_104}. These methods refine target position estimates in real-time, enhancing sensing accuracy. Incorporating such strategies improves the robustness of ISAC systems in practical deployments.
Observe that both the problems (\ref{PD_OP:1}) and (\ref{GMR_OP:1}) are highly non-convex due to the intractable expression of the $P_\mathrm{D}$ and $f_\mathrm{GM}(\mathbf{V}_\mathrm{RF},\mathbf{V}_\mathrm{BB})$, and also due to the non-convex unity magnitude constraint on each element 
of the RF TPC $\mathbf{V}_\mathrm{RF}$. Furthermore, the optimization variables $\mathbf{V}_\mathrm{BB}$ and $\mathbf{V}_\mathrm{RF}$ are tightly coupled in the objective functions and the constraints, which exacerbates the challenges in both the problems. In the following sections, we propose efficient algorithms for solving both problems.


\section{HBF design based on PD-max}\label{sec:PD-max}
This section proposes an efficient iterative algorithm for solving the optimization problem $\mathcal{P}_1$, given by (\ref{PD_OP:1}). 
Let us introduce a positive real slack variable $\eta \geq 0$ to make the problem $\mathcal{P}_1$ more flexible, so that its equivalent problem is given by
\begin{subequations}\label{PD_OP:2}
\begin{align} 
&\mathop {\max }\limits_{\mathbf{V}_\mathrm{RF}, ~\mathbf{V}_\mathrm{BB},~ \eta}   \eta  \\
&\text {s. t.} \quad P_\mathrm{D} \geq \eta,\label{cons:PD_EC} \\ &\quad \text{(\ref{cons:SINR}), (\ref{cons:TP}), and (\ref{cons:RF})}.
\end{align}
\end{subequations}
To solve the above problem (\ref{PD_OP:2}), we propose a bisection-Alt (Bi-Alt) algorithm in which the BB TPC $\mathbf{V}_\mathrm{BB}$ and RF TPC $\mathbf{V}_\mathrm{RF}$ are alternately optimized using the BCD technique. Subsequently, we update $\eta$ by employing the bisection search method. 
For a given $\eta \geq 0$, the equivalent feasible problem for  (\ref{PD_OP:2}) can be stated as
\begin{subequations}\label{PD_OP:4}
\begin{align}
& \min \limits_{\mathbf{V}_\mathrm{RF}, \mathbf{V}_\mathrm{BB}} \mathcal{J}\left(\mathbf{V}_\mathrm{RF}, \mathbf{V}_\mathrm{BB}\right) = \|\mathbf{V}_\mathrm{RF}\mathbf{V}_\mathrm{BB}\|_F^2 \\
&\mathrm{s.} ~\mathrm{t.}\quad \quad \text{(\ref{cons:PD_EC}), (\ref{cons:SINR}), and (\ref{cons:RF})}.  
\end{align}
\end{subequations}
Denoting the optimal solution of (\ref{PD_OP:4}) as \{$\mathbf{V}^*_\mathrm{RF}, \mathbf{V}^*_\mathrm{BB}$\}, it may be readily seen that if 
$\|\mathbf{V}^*_\mathrm{RF}\mathbf{V}^*_\mathrm{BB}\|_F^2 \leq P_\mathrm{t}$, problem (\ref{PD_OP:4}) is feasible, and infeasible otherwise. 
However, the problem (\ref{PD_OP:4}) above is still challenging to solve due to the intractable expression of $P_\mathrm{D}$ in (\ref{cons:PD_EC}) and owing to the non-convex constraints (\ref{cons:SINR}) and (\ref{cons:RF}). 
To overcome this, we first transform the intractable expression of $P_\mathrm{D}$ to a tractable form by exploiting its monotonicity with respect to the noncentrality parameter $\rho$, as demonstrated in Theorem 1 of \cite{sun2010monotonicity}. 
Consequently, the equivalent modified optimization problem of (\ref{PD_OP:4}) can be recast as follows
\begin{subequations}\label{PD_OP:5}
\begin{align} 
\hspace{4mm}& \quad \min \limits_{\mathbf{V}_\mathrm{RF}, \mathbf{V}_\mathrm{BB}} \mathcal{J}\left(\mathbf{V}_\mathrm{RF}, \mathbf{V}_\mathrm{BB}\right)  \\
&\mathrm{s.\hspace{1mm} t.} \quad \rho = \mu\left(\sum_{k=1}^{K} \left| \mathbf{a}^{H}_\mathrm{BS}({\theta})\mathbf{V}_{\rm RF}\mathbf{v}_{\mathrm{BB}, k} \right|^2\right)^2 \geq \widetilde{\eta},\label{cons:Pd_UP}\\ 
&\quad \quad \quad \quad \text{ (\ref{cons:SINR}) and (\ref{cons:RF})},
\end{align}
\end{subequations}
where $\widetilde{\eta}$ is the solution obtained for the corresponding value of $\rho$ by solving the following equation
\begin{equation} \label{eqn:MIN_NCP} 
1 - \mathcal {F}_{\chi _{2}^{2}(\rho)} \Big (\mathcal {F}_{\chi _{2}^{2}}^{-1}(1 - P_{\text {FA}}) \Big) = \eta.
\end{equation}
Although, the constraint (\ref{cons:Pd_UP}) is now tractable, the variables $\mathbf{V}_\mathrm{BB}$ and $\mathbf{V}_\mathrm{RF}$ are still tightly coupled in both the objective function and the constraint of (\ref{cons:Pd_UP}). To address this issue, we adopt the alternating optimization technique, where $\mathbf{V}_\mathrm{BB}$ and $\mathbf{V}_\mathrm{RF}$ are optimized in an alternating fashion until the objective function of (\ref{PD_OP:5}) $\mathcal{J}\left(\mathbf{V}_\mathrm{RF}, \mathbf{V}_\mathrm{BB}\right)$ converges. This is described as follows.
\subsection{Optimization of $\mathbf{V}_\mathrm{BB}$ for a fixed $\mathbf{V}_\mathrm{RF}$}\label{PD-BB}
For a fixed $\mathbf{V}_\mathrm{RF}$, the resultant optimization problem for $\mathbf{V}_\mathrm{BB}$ corresponding to (\ref{PD_OP:5}) is formulated as
\begin{subequations}\label{PD_OP:6}
\begin{align}
& \min \limits_{\mathbf{V}_\mathrm{BB}} \hspace{4mm} \sum_{k=1}^{K}\mathbf{v}_{\mathrm{BB},k}^{H}\mathbf{M} \mathbf{v}_{\mathrm{BB},k} \\
&\mathrm{s.} ~\mathrm{t.} \quad \sum\limits_{i\neq k}^{K}{\left\vert\mathbf{g}_k^H \mathbf{v}_{\mathrm{B B}, i}\right\vert^2} + \sigma_{k}^{2} \geq \frac{1}{\Gamma_k}\left\vert\mathbf{g}_k^H\mathbf{v}_{\mathrm{BB}, k}\right\vert^2,\forall k\label{cons:SINR_BB}\\ 
&\quad \quad \quad \sum_{k=1}^{K}\mathbf{v}_{\mathrm{BB},k}^{H}\mathbf{\Omega} \mathbf{v}_{\mathrm{BB},k} \geq \omega, \label{cons:Pd_BB}  
\end{align}
\end{subequations}
where $\mathbf{M} = \mathbf{V}^{H}_\mathrm{RF}\mathbf{V}_\mathrm{RF}\in \mathbb{C}^{N_\mathrm{RF} \times N_\mathrm{RF}}$, $\mathbf{g}_k = \mathbf{V}^{H}_\mathrm{RF}\mathbf{h}_k \in \mathbb{C}^{N_\mathrm{RF} \times 1}$, $\mathbf{\Omega} = \mathbf{V}^{H}_\mathrm{RF}{\mathbf{A}}(\theta)\mathbf{V}_\mathrm{RF}\in \mathbb{C}^{N_\mathrm{RF} \times N_\mathrm{RF}}$, and $\omega = \sqrt{\frac{\widetilde{\eta}}{\mu}}$. 
Observe that the problem (\ref{PD_OP:6}) is non-convex due to the non-convex constraints (\ref{cons:SINR_BB}) and (\ref{cons:Pd_BB}). In order to handle the non-convexity of constraints (\ref{cons:SINR_BB}), we apply a common phase shift to $\mathbf{v}_{\mathrm{BB},k}, \forall k,$ so that the quantity $\left\vert\mathbf{g}_k^H\mathbf{v}_{\mathrm{BB}, k}\right\vert$ leads to real-valued quantities $\forall k$. Consequently, the constraints (\ref{cons:SINR_BB}) are transformed into the second-order cone (SOC) convex constraints as follows
\begin{equation}\label{eqn:SOCP}
\begin{aligned}
\left \|{ \begin{array}{c}  \mathbf{G}^H \mathbf{e}\\ \sigma_{k} \end{array} }\right \|_2 \le  \sqrt{1+\frac{1}{\Gamma_k}} t_{k, i}, ~\forall k,
\end{aligned}
\end{equation}
where $\mathbf{G}^H = [\mathbf{g}_1, \hdots, \mathbf{g}_1] \in \mathbb{C}^{N_\mathrm{RF}\times K}$ and $t_{k, i}= \mathbf{g}_k^H \mathbf{v}_{\mathrm{BB}, i}$. Furthermore, the quantity  $\mathbf{e}\in \mathbb{C}^{K \times 1}$ is the elementary vector with one on its $k$th position and zero elsewhere. 
Next, to handle the non-convex constraint (\ref{cons:Pd_BB}), we employ the popular successive convex approximation (SCA) technique, where the first-order Taylor series is used to find the pertinent convex surrogate functions for (\ref{cons:Pd_BB}). Specifically, a convex lower bound in a linear form of the function $\mathbf{v}_{\mathrm{BB}, k}^{H}\mathbf{\Omega} \mathbf{v}_{\mathrm{BB}, k}$ around the point $\mathbf{{v}}^{(j)}_{\mathrm{BB},k}$ can be constructed as follows
\begin{equation}\label{eqn:SFD} 
\mathbf{v}_{\mathrm{BB}, k}^H\mathbf{\Omega}\mathbf{v}_{\mathrm{BB}, k} \geq 2\Re\{\mathbf{{v}}_{\mathrm{BB},k}^{(j)H}\mathbf{\Omega}\mathbf{v}_{\mathrm{BB},k}\}-\mathbf{{v}}_{\mathrm{BB}, k}^{(j)H}\mathbf{\Omega}\mathbf{{v}}_{\mathrm{BB},k}^{(j)},
\end{equation}
where $\mathbf{v}^{(j)}_{\mathrm{BB},k}$, $\forall k$, denotes the optimal solution obtained at the $j$th iteration. 
Following these transformations, the equivalent convex problem for (\ref{PD_OP:6}) at the $(j+1)$th iteration can be formulated as 
\begin{subequations}\label{PD_OP:7}
\begin{align}
& \min \limits _{\mathbf{V}_\mathrm{BB}} \hspace{4mm} \sum_{k=1}^{K}\mathbf{v}_{\mathrm{BB}, k}^{H}\mathbf{M} \mathbf{v}_{\mathrm{BB}, k} \\
&\mathrm{s.} ~\mathrm{t.} \quad 2\sum_{k=1}^{K}\Re\{\mathbf{v}_{\mathrm{BB}, k}^{(j)H}\mathbf{\Omega}\mathbf{v}_{\mathrm{BB}, k}\} \geq \widetilde{\omega},\text{ and (\ref{eqn:SOCP})}, \label{cons:PD_RC}
\end{align}
\end{subequations}
where $\widetilde{\omega} = \omega + \sum_{k=1}^{K}\mathbf{v}_{\mathrm{BB}, k}^{(j)H}\mathbf{\Omega}\mathbf{v}_{\mathrm{BB}, k}^{(j)}$. Since the problem (\ref{PD_OP:7}) is convex, it can be efficiently solved via a standard convex optimization tool. Thereby, we follow the SCA approach for recursively solving the problem (\ref{PD_OP:7}), which obtains the optimal solution to the original problem (\ref{PD_OP:6}). Moreover, we show the key steps involved in optimizing $\mathbf{V}_\mathrm{BB}$ in Algorithm \ref{alg:SCA}.
\begin{algorithm}[t]
\caption{SCA method for solving (\ref{PD_OP:6})}
 \textbf{Input:} $\mathbf{V}_\mathrm{RF}$, $\mathbf{V}_\mathrm{BB}$ \\ 
 \textbf{Output:} optimal BB TPC $\mathbf{V}^*_\mathrm{BB}$
 \label{alg:SCA}
\begin{algorithmic}[1]
\State \textbf{initialize:} $j=0$, $\mathbf{{v}}^{(j)}_{\mathrm{BB},k} = \mathbf{V}_\mathrm{BB}(:,k)$, $\forall k$ 
\Repeat 
     \State compute $\widetilde{\omega}$ as $\widetilde{\omega} = \omega + \sum_{k=1}^{K}\mathbf{v}_{\mathrm{BB}, k}^{(j)H}\mathbf{\Omega}\mathbf{v}_{\mathrm{BB}, k}^{(j)}$
     \State update $\mathbf{v}^{(j+1)}_{\mathrm{BB}, k}$, $\forall k$ by solving (\ref{PD_OP:7})
     \State set $ j\leftarrow j+1$
 \Until objective function of (\ref{PD_OP:7}) converges
\end{algorithmic}
\end{algorithm}

\subsection{Optimization of $\mathbf{V}_\mathrm{RF}$ for a fixed $\mathbf{V}_\mathrm{BB}$} \label{PD-RF}
In this subsection, we optimize $\mathbf{V}_\mathrm{RF}$ for a fixed $\mathbf{V}_\mathrm{BB}$. 
To this end, let us define $\mathbf{H}_k=\mathbf{h}_k\mathbf{h}^H_k \in \mathbb{C}^{N_\mathrm{t}\times N_\mathrm{t}}$, $\mathbf{B}_i=\mathbf{v}_{\mathrm{BB}, i}\mathbf{v}^H_{\mathrm{BB},i}\in \mathbb{C}^{N_\mathrm{RF}\times N_\mathrm{RF}}$, and $\Delta_k \triangleq  \Gamma_k{\sum \limits_{\begin{subarray}{l}  i \ne k \end{subarray}}^{K}  \mathrm{tr}(\mathbf{V}_{\rm RF}^H\mathbf{H}_i\mathbf{V}_\mathrm{RF}\mathbf{B}_i)} - \mathrm{tr}\left({\mathbf{V}^H_\mathrm{RF}\mathbf{H}_k\mathbf{V}_\mathrm{RF}\mathbf{B}_k}\right) + \sigma^2_{k}\Gamma_k$ with $\mathbf{B} =\mathbf{V}_{\mathrm{ BB}}\mathbf{V}^H_{\mathrm{BB}}\in \mathbb{C}^{N_\mathrm{RF} \times N_\mathrm{RF}}$.
Thus, we formulate the sub-problem of optimizing $\mathbf{V}_\mathrm{RF}$ as follows
\begin{subequations}\label{PD_OP:8}
\begin{align}
& \min \limits _{\mathbf{V}_\mathrm{RF}} \hspace{4mm} {\mathrm{tr}\left({\mathbf{V}_{\rm RF}^H\mathbf{I}_{N_{t}}\mathbf{V}_{\rm RF}\mathbf{B}}\right)}\label{OB_RF} \\
&\mathrm{s.} ~\mathrm{t.} \quad \Delta_{k} \leq 0 ,\forall k, \label{cons:SINR_RF}\\ 
&\quad \quad {\mathrm{tr}\left({\mathbf{V}_{\rm RF}^H{\mathbf{A}}(\theta)\mathbf{V}_{\rm RF}\mathbf{B}}\right)}\geq \omega.\label{cons:Pd_RF} 
\end{align}
\end{subequations}
By leveraging the matrix transformation identity $\mathrm{tr}\left({\mathbf{P}^H\mathbf{Q}\mathbf{P}\mathbf{R}}\right) = \text{vec}\ (\mathbf {P})^{H} \left ({\mathbf {R}^{T} \otimes \mathbf {Q}}\right)\text{vec}\ (\mathbf {P})$, one can rewrite the objective function (\ref{OB_RF}), and constraints (\ref{cons:SINR_RF}) and (\ref{cons:Pd_RF}) as given by (\ref{jit_1}), (\ref{jit_2}), and (\ref{jit_3}), respectively, shown at the top of the next page.
\begin{figure*}[t] 
\begin{align}
\mathrm{tr}\left({\mathbf{V}_{\rm RF}^H\mathbf{I}_{N_{t}}\mathbf{V}_{\rm RF}\mathbf{B}}\right)& = \text{vec}\ (\mathbf{V}_{\text{RF}})^{H}(\mathbf{B}^{T}\otimes \mathbf{I}_{N_{t}})\ \text{vec}\ (\mathbf{V}_{\text{RF}}) \label{jit_1}\\
\Delta_k& = \delta\sum \limits_{\begin{subarray}{l}  i \ne k \end{subarray}}^{K}\text{vec}\ (\mathbf{V}_{\text{RF}})^{H} \left(\mathbf{B}_i^{T} \otimes \mathbf{H}_k\right) \text{vec}\ (\mathbf{V}_{\text{RF}}) -\text{vec}\ (\mathbf{V}_{\text{RF}})^{H} \left(\mathbf{B}_k^{T} \otimes \mathbf{H}_k\right) \mathrm{vec}(\mathbf{V}_{\mathrm{RF}}) + \sigma_{k}^{2}\delta \leq 0,\forall k, \label{jit_2}\\
\mathrm{tr}\left({\mathbf{V}_{\rm RF}^H{\mathbf{A}}(\theta)\mathbf{V}_{\rm RF}\mathbf{B}}\right)& = \text{vec}\ (\mathbf{V}_{\text{RF}})^{H}(\mathbf{B}^{T}\otimes {\mathbf{A}}(\theta))\ \text{vec}\ (\mathbf{V}_{\text{RF}})\label{jit_3}
\end{align}
\normalsize
\hrulefill
\end{figure*}
Following these transformations, the problem (\ref{PD_OP:8}) can be recast as 
\begin{subequations}\label{PD_OP:9}
\begin{align}
\min \limits_{\boldsymbol{\phi}} \quad & \boldsymbol{\phi}^H \mathbf{T} \boldsymbol{\phi} \label{OF:jit}\\ \mathrm{s}.~\mathrm{t}.\ & \boldsymbol{\phi}^H \mathbf{\Pi}_k \boldsymbol{\phi} + \sigma_{k}^{2}\Gamma_k \leq 0 ,\forall k,\label{cons:vec_SINR_RF}\\ 
& \boldsymbol{\phi}^H \boldsymbol{\Lambda} \boldsymbol{\phi} \geq \omega,\label{cons:vec_PD_RF}\\ & \vert\boldsymbol{\phi}(n)\vert=1, \forall n \in \{1,\hdots,N_\mathrm{t}N_\mathrm{RF} \}, \label{cons:vec_UM} 
\end{align}
\end{subequations}
where $\boldsymbol{\phi} = \text{vec}\ (\mathbf{V}_{\text{RF}}) \in \mathbb{C}^{N_\mathrm{t}N_\mathrm{RF}\times 1}$, $\mathbf{T} = \left(\mathbf{B}^{T}\otimes \mathbf{I}_{N_{t}}\right) \in \mathbb{C}^{N_\mathrm{t}N_\mathrm{RF}\times N_\mathrm{t}N_\mathrm{RF}} $, $\boldsymbol{\Lambda} = \left(\mathbf{B}^{T}\otimes {\mathbf{A}}\left(\theta\right)\right) \in \mathbb{C}^{N_\mathrm{t}N_\mathrm{RF}\times N_\mathrm{t}N_\mathrm{RF}}$, and  $\mathbf{\Pi}_k = \Gamma_k \Big (\sum \limits_{\begin{subarray}{l}  i \ne k \end{subarray}}^{K} \left(\mathbf{B}_i^{T} \otimes \mathbf{H}_k\right) \Big) - \left(\mathbf{B}_k^{T} \otimes \mathbf{H}_k \right) \in \mathbb{C}^{N_\mathrm{t}N_\mathrm{RF}\times N_\mathrm{t}N_\mathrm{RF}}$. 
Furthermore, the problem (\ref{PD_OP:9}) above is still challenging to solve in the Euclidean space due to the non-convex unity magnitude constraints (\ref{cons:vec_UM}). To solve this problem, we propose a penalty-based RCG (PRCG) algorithm that solves the problem in the Riemannian space. 
Specifically, we first add the constraints (\ref{cons:vec_SINR_RF}) and (\ref{cons:vec_PD_RF}) to the objective function as penalty terms. Subsequently, we employ the RCG framework to tackle the unity magnitude constraint (\ref{cons:vec_UM}), which solves the problem on the complex circle Riemannian manifold. Finally, we adjust the penalty term to satisfy the constraints (\ref{cons:vec_SINR_RF}) and (\ref{cons:vec_PD_RF}).
\begin{algorithm}[t]
\caption{PRCG algorithm for solving (\ref{PD_OP:8})}
 \textbf{Input:} $\mathbf{V}_\mathrm{RF} \in \mathcal{M}$, $\mathbf{V}_\mathrm{BB}$, $\lambda \geq 1$, $0< c < 1$, and thresholds $\epsilon_1>0, \epsilon_2 >0$\\
 \textbf{Output:}  Optimal $\mathbf{V}^*_\mathrm{RF}$
 \label{alg:EPMO}
\begin{algorithmic}[1]
\State \textbf{initialize:} $t = 0$, $\begin{aligned}[t]
    \boldsymbol{\phi}_t &= \text{vec}(\mathbf{V}_\mathrm{RF}),
    \zeta_t = -\nabla_{\mathcal{M}} \mathcal{L}(\boldsymbol{\phi}_t)
\end{aligned}$
\While{$\left(\|\nabla _{\mathcal {M}} \mathcal{L}({\boldsymbol{\phi}_t}) \|_2^2 \geq \epsilon_1 \right)$}
    \State choose step size $\nu_2$ using Armijo backtracking line search algorithm 
    \State update the next point $\boldsymbol{\phi}_{t+1}$ by using (\ref{Ret})
    \State evaluate $\nabla_{\mathcal{M}} \mathcal{L}\left({\boldsymbol{\phi}_{t+1}}\right)$ using (\ref{eqn:RG}.
    \State compute $\boldsymbol{\zeta}_{t+1}$ according to (\ref{SD})
    \State set $t\leftarrow t+1$
    \EndWhile
    \State \textbf{end} \textbf{while}
    \State $\textbf{if} ~\big(\psi_k\left(\boldsymbol{\phi}_{t}\right) \leq \epsilon_2~ \& \& ~\delta\left(\boldsymbol{\phi}_t\right) \leq \epsilon_2 \big), \forall k $
    \State \hspace{0.5cm} \textbf{return} $\boldsymbol{\phi}^{*} = \boldsymbol{\phi}_t $ \textbf{stop}
    \State $\textbf{else}$
    \State \hspace{0.5cm} update $\mathbf{\lambda} = \frac{\lambda}{c}$ and go to step 2
    \State $\textbf{end}$ $\textbf{if}$
    \State \textbf{return:} restore matrix $\mathbf{V}^*_\mathrm{RF}$ via $\boldsymbol{\phi}^{*}$ 
\end{algorithmic}
\end{algorithm}

To this end, let us convert the constrained problem (\ref{PD_OP:9}) into an unconstrained problem on the Riemannian manifold upon adding the constraints (\ref{cons:vec_SINR_RF}) and  (\ref{cons:vec_PD_RF}) into the objective function as a penalty term. Consequently, the resulting problem is formulated as
\begin{equation}\label{RCG_1}
\begin{aligned}
& \min_{\boldsymbol{\phi}}~  \mathcal{L}(\boldsymbol{\phi}) = \boldsymbol{\phi}^H \mathbf{T} \boldsymbol{\phi} + \lambda \left(\sum_{\begin{subarray}{l} k = 1 \\  \end{subarray}}^{K} \psi_k\left(\boldsymbol{\phi}\right)+\delta\left(\boldsymbol{\phi}\right)\right)\\ 
&\text{s.} ~ \text{t.} \quad (\ref{cons:vec_UM}),
\end{aligned}
\end{equation}
where $\lambda \geq 1$ is a penalty parameter, which maintains a trade-off between the objective function (\ref{OF:jit}), and the constraints (\ref{cons:vec_SINR_RF}) and (\ref{cons:vec_PD_RF}). Furthermore, the quantities $\psi_k(\boldsymbol{\phi})$ and $\delta(\boldsymbol{\phi})$ are defined as follows
\begin{subequations} \label{eqn:jit_4}
\begin{align}
 \psi_k(\boldsymbol{\phi}) \triangleq &\Big(\max\big\{0, \boldsymbol{\phi}^H \mathbf{\Pi}_k \boldsymbol{\phi} + \sigma_{k}^{2}\Gamma_k\big\}\Big)^2,\ \forall k \\
 \delta(\boldsymbol{\phi}) \triangleq &\Big(\max\big\{0, \omega - \boldsymbol{\phi}^H \boldsymbol{\Lambda} \boldsymbol{\phi} \big\}\Big)^2.
\end{align}
\end{subequations}
Clearly, for the given $\lambda$, the problem (\ref{eqn:jit_4}) above represents a manifold optimization problem, where the non-convex unity magnitude constraint (\ref{cons:vec_UM}) forms a $N_\mathrm{t}N_\mathrm{RF}$-dimensional complex circle Riemannian manifold, as $\mathcal{M} = \{\boldsymbol{\phi} \in {\mathbb C}^{{N_\mathrm{t}}{N_\mathrm{RF}} \times 1}:  \left|\boldsymbol{\phi}(n)\right| = 1,\forall   1 \leq n \leq N_\mathrm{t}N_\mathrm{RF}\}$. 
Thus, we adopt the RCG method to solve (\ref{eqn:jit_4}), which differs from the conventional gradient descent method performed in the Euclidean space. The following are the key steps involved in each iteration of the PRCG method.
\subsubsection*{Riemannian gradient}
We evaluate the Riemannian gradient at point $\boldsymbol{\phi}$ denoted as $\nabla _{\mathcal{M}} \mathcal{L}(\boldsymbol{\phi})$ by orthogonally projecting the Euclidean gradient onto the tangent space of the manifold $T_{\boldsymbol{\phi}} \mathcal {M}$ expressed as
\begin{equation}
 T_{\boldsymbol{\phi}} \mathcal {M} = \lbrace \mathbf {z}\in {{\mathbb {C}}^{N_{t}N_{RF}\times 1}}|\Re\{\mathbf {z} \odot \boldsymbol{\phi}^*\} = \mathbf {0}_{N_{t}N_{RF}\times 1}\rbrace.  
 \end{equation}
Thus, the Riemannian gradient is given by
\begin{subequations}\label{eqn:RG} 
\begin{align}
\nabla _{\mathcal {M}} \mathcal{L}(\boldsymbol{\phi}) & = {\text{Pro}}{{\text{j}}_{\boldsymbol{\phi}}}\nabla \mathcal{L}(\boldsymbol{\phi}) \\ 
& = \nabla \mathcal{L}(\boldsymbol{\phi}) - \Re\{ \nabla\mathcal{L}(\boldsymbol{\phi}) \odot {{\boldsymbol{\phi}}^{*}}\} \odot \boldsymbol{\phi}, 
 \end{align} 
\end{subequations}
where $\text{Proj}$ denotes the projection operation, and $\nabla \mathcal{L}(\boldsymbol{\phi})$ is the Euclidean gradient. Furthermore, we evaluate $\nabla \mathcal{L}(\boldsymbol{\phi})$ as follows
\begin{equation}\label{eqn:EQN_GR}  
\nabla \mathcal{L}(\boldsymbol{\phi}) =2 \mathbf{T} \boldsymbol{\phi} +  \lambda \left( \sum_{k=1}^K \boldsymbol{\varkappa}_k + \boldsymbol{\varsigma}\right),  
\end{equation}

where the quantities $\boldsymbol{\varkappa}_k$ and $ \boldsymbol{\varsigma}$ are given by
\begin{subequations}\label{eqn:EG_SR} 
\begin{align}
&\boldsymbol{\varkappa}_k = \begin{cases} 4\left(\boldsymbol{\phi}^H \mathbf{\Pi}_k \boldsymbol{\phi} + \sigma_{k}^{2}\Gamma_k\right) \boldsymbol{\Pi}_k \boldsymbol{\phi}, & \boldsymbol{\phi}^H \mathbf{\Pi}_k \boldsymbol{\phi} + \sigma_{k}^{2}\Gamma_k \geq 0 \\ \mathbf{0}, & \boldsymbol{\phi}^H \mathbf{\Pi}_k \boldsymbol{\phi} + \sigma_{k}^{2}\Gamma_k < 0\end{cases}
\\
& \boldsymbol{\varsigma} = \begin{cases} 4\left(\boldsymbol{\phi}^H \boldsymbol{\Lambda}\boldsymbol{\phi}-\omega\right)\boldsymbol{\Lambda}\boldsymbol{\phi}, & \omega \geq \boldsymbol{\phi}^H \boldsymbol{\Lambda} \boldsymbol{\phi}  \\ 0, & \omega < \boldsymbol{\phi}^H \boldsymbol{\Lambda} \boldsymbol{\phi}\end{cases}.
\end{align}
\end{subequations}


\subsubsection*{Steepest search direction}
With the aid of the Riemannian gradient, we find the most efficient steepest search direction $\boldsymbol{\zeta}_{j+1}$ at the $(j+1)$th iteration by employing the conjugate gradient method as follows
\begin{equation} \label{SD}
\begin{aligned}
\boldsymbol{\zeta }^{(j+1)} = -\nabla _{\mathcal {M}} \mathcal{L}\left(\boldsymbol{\phi}^{(j+1)}\right) + \nu_1\; T_{\boldsymbol{\phi}^(j) \mapsto \boldsymbol{\phi}^{(j+1)}} \left(\boldsymbol{\zeta }^{(j)} \right), 
\end{aligned} 
\end{equation}
where $\nu_1$ represents the Polak-Ribiére’s conjugate parameter \cite{jite_1} and $T_{\boldsymbol{\phi}^{(j)} \mapsto \boldsymbol{\phi}^{(j+1)}} \left(\boldsymbol{\zeta }^{(j)} \right)$ is the transport operation. Note that the vectors $\boldsymbol{\zeta }^{(j+1)}$ and $\boldsymbol{\zeta }^{(j)}$ cannot be added directly, since they belong to different tangent spaces. Thereby, we require a transport operation $T_{\boldsymbol{\phi}^{(j)} \mapsto \boldsymbol{\phi}^{(j+1)}} \left(\boldsymbol{\zeta}^{(j)} \right)$ to map the search direction from its original tangent space to the current tangent space, which is defined as
\begin{equation}\label{TV}
{\mathcal{T}}_{{\boldsymbol{\phi}^{(j)}}\mapsto {\boldsymbol{\phi}}^{(j+1)}} \left({\mathbf {\zeta}}^{(j)} \right) 
= \mathbf{\zeta}^{(j)} - \Re \left\lbrace \mathbf {\zeta}^{(j)} \odot \left(\boldsymbol{\phi}^{(j+1)}\right)^* \right\rbrace 
\odot \boldsymbol{\phi}^{(j+1)}.
\end{equation} 

\subsubsection*{Retraction}
Similar to the classic gradient descent approach, we compute the next point $\boldsymbol{\phi}^{(j+1)}$ using the step size $\nu_2$ and search direction $\boldsymbol{\zeta}^{(j)}$. Yet, it is very likely that the next point $\boldsymbol{\phi}^{(j+1)} = \boldsymbol{\phi}^{(j)} + \nu_2\boldsymbol{\zeta}^{(j)}$ does not fall on the manifold $\mathcal{M}$, but rather lies on the $T_{\boldsymbol{\phi}^{(j+1)}} \mathcal {M}$. Therefore, to retract back the point on the manifold $\mathcal {M}$, we perform the retraction mapping as follows
\begin{equation}\label{Ret}
\begin{aligned}
& {\text{Ret}}{{\text{r}}_{\boldsymbol{\phi}}}:{\mathcal{T}_{\boldsymbol{\phi}}}\mathcal{M} \to \mathcal{M}: \\ 
& \boldsymbol{\phi}^{(j+1)} = \left[ {\frac{{{{({\boldsymbol{\phi}^{(j)} +}\nu_2\boldsymbol\zeta^{(j)})_1}}}}{{|{{({\boldsymbol{\phi}^{(j)} +}\nu_2\boldsymbol\zeta^{(j)})_1}}|}}},\hdots,{\frac{{{{({\boldsymbol{\phi}^{(j)} +}\nu_2\boldsymbol\zeta^{(j)})_{N_\mathrm{t}N_\mathrm{RF}}}}}}{{|{{({\boldsymbol{\phi}^{(j)} +}\nu_2\boldsymbol\zeta^{(j)})_{N_\mathrm{t}N_\mathrm{RF}}}}|}}}\right]^T, 
\end{aligned}
\end{equation} 
where the step size $\nu_2$ is obtained by the Armijo backtracking line search algorithm \cite{jite_1}. 
\subsubsection*{Update penalty parameter}
Finally, we update the penalty parameter $\lambda$ to meet the constraints (\ref{cons:vec_SINR_RF}) and (\ref{cons:vec_PD_RF}) in the optimization problem (\ref{PD_OP:9}).
Note that the penalty parameter $\lambda$ is crucial for achieving an optimal feasible $\mathbf{V}_{\mathrm{RF}}$. If $\lambda$ is too small, the resultant solution may fall far outside the feasible region, potentially leading to the violation of the constraints. Thus, we initialize $\lambda$ to be a small number and then gradually increase it as $\lambda = \lambda/c$, where $c \in (0, 1)$ is the scaling parameter for ensuring that the constraint violations are sufficiently penalized. 
We summarize the PRCG method in Algorithm \ref{alg:EPMO}, which is guaranteed to converge to a stationary point \cite{jite_2}.
\subsection{Slack variable $\eta$ update}
Finally, we update the slack variable $\eta$ via the bisection search method for fixed values of $\mathbf{V}_\mathrm{RF}$ and $\mathbf{V}_\mathrm{BB}$. Toward this, we define the quantities $\eta_{L}$ and $\eta_U$ as the lower and upper bound of the optimal value of $P_\mathrm{D}$, respectively. 
Since the probability of detection is bounded between 0 and 1, we set 
$\eta_L = 0$ and $\eta_U = 1$ to keep the optimization within a valid range, ensuring accurate constraint handling.
When the problem $(\ref{PD_OP:4})$ is feasible for a given $\eta$ i.e., $\|\mathbf{V}^*_\mathrm{RF}\mathbf{V}^*_\mathrm{BB}\|_F^2 \leq P_\mathrm{t}$, we update the lower bound as $\eta_{L} = \eta$, else we update the upper bound as $\eta_U=\eta$.
\begin{algorithm}[t]
\caption{Bi-Alt method for solving $\mathcal{P}_1$}
 \textbf{Input:}  $P_\mathrm{t}$, $P_\mathrm{FA}$, $\Gamma_k,\forall k$, thresholds $\epsilon_3$, $\epsilon_4$
\begin{algorithmic}[1]\label{alg:Bi_BCD}
\State \textbf{initialize:} $\mathbf{V}_\mathrm{RF},\mathbf{V}_\mathrm{BB}$, lower bound $\eta_\mathrm{L} = 0$, and upper bound $\eta_\mathrm{U} = 1$.
         \Repeat
                \State $\eta = \left(\eta_\mathrm{L} + \eta_\mathrm{U}\right) / 2$
                \State compute $\widetilde{\eta}$ using (\ref{eqn:MIN_NCP})
                \Repeat 
                \State set $ j = 0 $, $\mathcal{J}^{(j)} = \infty$.                       
                \State evaluate $\mathbf{V}^{(j + 1)}_\mathrm{BB}$ for given $\mathbf{V}^{(j )}_\mathrm{RF}$ by solving (\ref{PD_OP:6}) via Algorithm \ref{alg:SCA}.
                \State obtain $\mathbf{V}^{(j + 1)}_\mathrm{RF}$ for given  $\mathbf{V}^{(j + 1)}_\mathrm{BB}$ by solving (\ref{PD_OP:8}) via Algorithm \ref{alg:EPMO}.
                \State compute $\mathcal{J}^{(j+1)} = \|\mathbf{V}^{(j+1)}_\mathrm{RF}\mathbf{V}^{(j+1)}_\mathrm{BB}\|^2_F$
                \State set $j\leftarrow j+1$
                \Until $\big\vert({\mathcal{J}^{(j)} - \mathcal{J}^{(j - 1)}} \big\vert \leq \epsilon_3 $
                \State \textbf{if} obtained set $\{\mathbf{V}_\mathrm{RF}, \mathbf{V}_\mathrm{BB}\}$ is feasible,
                \State update $\eta_\mathrm{L} = \eta$  
                \State \textbf{else} 
                \State set $\eta_\mathrm{U} = \eta$.
           \Until $ \eta_\mathrm{U} - \eta_\mathrm{L}  \leq \epsilon_4$    
\end{algorithmic}
\end{algorithm}


Based on the method presented above, we summarize the complete Bi-Alt method conceived for solving the PD-max problem in Algorithm 3. Note that $\epsilon_3$ and $\epsilon_4$ in Algorithm 3 are the error tolerance for the function $\mathcal{J}\left(\mathbf{V}_\mathrm{RF}, \mathbf{V}_\mathrm{BB}\right)$ in the inner loop and bisection search in the outer loop, respectively. 
Furthermore, the computational complexity of Algorithm 3 depends on Algorithm \ref{alg:SCA} and \ref{alg:EPMO}. Since Algorithm \ref{alg:SCA} involves the interior-point method harnessed for optimizing the BB TPC $\mathbf{V}_\mathrm{BB}$, it has the complexity order of $\mathcal{O}(\mathcal{I}_\mathrm{b}N^{3.5}_\mathrm{RF}K^{3.5})$, where $\mathcal{I}_\mathrm{b}$ denotes the number of iterations required for updating $\mathbf{V}_\mathrm{BB}$. 
On the other hand, Algorithm \ref{alg:EPMO} focuses on optimizing the RF TPC $\mathbf{V}_\mathrm{RF}$, whose main complexity arises from the computation of the Euclidean gradient (\ref{eqn:EQN_GR}). Consequently, the complexity of Algorithm \ref{alg:EPMO} is given by $\mathcal{O}\left(\mathcal{I}_\mathrm{r}N^2_\mathrm{t}N^2_\mathrm{RF}\frac{1}{\epsilon^2_1}\right)$, where $\mathcal{I}_\mathrm{r}$ represents the combined number of iterations required for the RCG method and updating the penalty factor $\lambda$. Therefore, the overall computational complexity of Algorithm 3 is given by $\mathcal{O}\left(\mathcal{I}_\mathrm{out}\mathcal{I}_\mathrm{in}\left(\mathcal{I}_\mathrm{b}N^{3.5}_{RF}K^{3.5}\log_2 \left( \frac{1}{\epsilon_0} \right) + \mathcal{I}_\mathrm{r}N^2_{t}N^2_{RF}\frac{1}{\epsilon^2_1}\right)\right)$, where $\mathcal{I}_\mathrm{in}$ and $\mathcal{I}_\mathrm{out}$ denote the number of iterations required in the inner layer and outer layer, respectively. In addition, $\epsilon_0$ signifies the accuracy of the SCA in Algorithm \ref{alg:SCA}.

\section{HBF design based on GMR-max}
In this section, we focus our attention on optimizing the BB and RF TPCs based on the GMR-max problem given by $\mathcal{P}_2$ of (\ref{GMR_OP:1}), which is NP-hard. To solve $\mathcal{P}_2$, we first utilize the relationship between the PD $P_\mathrm{D}$ and the noncentrality parameter $\rho$, as discussed in Section \ref{sec:PD-max}, which transforms the constraint (\ref{cons:Pd}) into a tractable form. 
Given the transformed PD constraint (\ref{cons:Pd}), the equivalent GMR-max problem is reformulated as
\begin{subequations}\label{GM_OP:2}
\begin{align} 
 \hspace{4mm}&\mathop {\max }\limits_{\mathbf{V}_\mathrm{RF}, \mathbf{V}_\mathrm{BB}} \quad f_\mathrm{GM}(\mathbf{V}_\mathrm{RF},\mathbf{V}_\mathrm{BB})   \label{GM_OF:2}\\
&\mathrm{s.\hspace{1mm} t.} \text{(\ref{cons:TP}), (\ref{cons:RF}), and (\ref{cons:Pd_UP})}.
\end{align}
\end{subequations}
Observe that the objective function $f_\mathrm{GM}(\mathbf{V}_\mathrm{RF}, \mathbf{V}_\mathrm{BB})$ in the above problem (\ref{GM_OP:2}) is nonlinear and involves the product of non-concave functions $R_k(\mathbf{V}_{\rm RF},\mathbf{V}_\mathrm{BB})$, which renders $f_\mathrm{GM}(\mathbf{V}_\mathrm{RF}, \mathbf{V}_\mathrm{BB})$ a highly non-convex function. To handle this hurdle, we first transform $f_\mathrm{GM}(\mathbf{V}_\mathrm{RF}, \mathbf{V}_\mathrm{BB})$ from a nonlinear function to a weighted linear function of the CU's rate.
Toward this, let us consider $(\mathbf{V}^{(0)}_\mathrm{RF}, \mathbf{V}^{(0)}_\mathrm{BB})$ as an initial feasible point, and $(\mathbf{V}^{(j)}_\mathrm{RF},\mathbf{V}^{(j)}_\mathrm{BB})$ as the feasible solution obtained from the $(j-1)$th iteration. Then, the linearized function of the composite function $f_\mathrm{GM}(\mathbf{V}_{\rm RF},\mathbf{V}_\mathrm{BB})$ around the point $R_k\left(\mathbf{V}^{(j)}_\mathrm{RF},\mathbf{V}^{(j)}_\mathrm{BB}\right)$ is given as \cite{Without_ISAC_3} 
\begin{equation} 
\frac{1}{K}f_\mathrm{GM}\left(\mathbf{V}^{(j)}_\mathrm{RF},\mathbf{V}^{(j)}_\mathrm{BB}\right)\left[\sum _{k=1}^{K}\frac{R_k(\mathbf{V}_{\rm RF},\mathbf{V}_\mathrm{BB})} {R_k\left(\mathbf{V}^{(j)}_\mathrm{RF},\mathbf{V}^{(j)}_\mathrm{BB}\right)}\right]. 
\end{equation}
Thus, the problem (\ref{GM_OP:2}) is equivalently transformed to the following weighted sum rate maximization problem at the $j$th iteration
\begin{subequations}\label{GM_OP:4}
\begin{align} 
&\mathop {\max}\limits_{\mathbf{V}_\mathrm{RF}, \mathbf{V}_\mathrm{BB}} \quad g^{(j)}(\mathbf{V}_\mathrm{RF},\mathbf{V}_\mathrm{BB}) \triangleq  \sum _{k=1}^{K} {u}^{(j)}_k R_k(\mathbf{V}_\mathrm{RF},\mathbf{V}_\mathrm{BB}) \label{GM_OF:4}\\
&\mathrm{s.\hspace{1mm} t.} \quad \text{(\ref{cons:TP}), (\ref{cons:RF}), and (\ref{cons:Pd_UP})},
\end{align}
\end{subequations}
where ${u}^{(j)}_k$ is the weight corresponding to CU $k$, which is computed as
\begin{equation}\label{eqn:AW}
u^{(j)}_k  =  \frac{\displaystyle \max _{k^{\prime }\in {\mathcal K}} R_{k^{\prime }}\left(\mathbf{V}^{(j)}_{\rm RF},\mathbf{V}^{(j)}_\mathrm{BB}\right)} {R_k\left(\mathbf{V}^{(j)}_\mathrm{ RF},\mathbf{V}^{(j)}_\mathrm{BB}\right)}, \ \forall k.
\end{equation}
Although, the objective function (\ref{GM_OF:2}) is transformed to the linearized form (\ref{GM_OF:4}), it is still non-convex due to the multiple fractional parameters of SINR terms $\mathbf{\gamma}_k(\mathbf{V}_\mathrm{RF},\mathbf{V}_\mathrm{BB})$. Moreover, the tightly coupled variables $\mathbf{V}_\mathrm{RF}$ and $\mathbf{V}_\mathrm{BB}$, both in the objective function and constraints, make the problem (\ref{GM_OP:4}) even more challenging to solve. To solve this problem, we propose a majorization and minimization-based alternating (MM-Alt) algorithm, in which for a fixed ${u}^{(j)}_k, \forall {k}$, at the $j$th iteration, we first split the problem (\ref{GM_OP:4}) into two sub-problems for the optimization of the BB TPC $\mathbf{V}_\mathrm{BB}$ and RF TPC $\mathbf{V}_\mathrm{RF}$ alternatively. Furthermore, at each stage of optimization of these variables, we transform the non-convex objective functions corresponding to (\ref{GM_OF:4}) into suitable convex surrogate functions via the MM technique \cite{Without_ISAC_2,zhang2021rate,zhang2023discerning}. 
\subsection{Optimization of $\mathbf{V}_\mathrm{BB}$ for a fixed $\mathbf{V}_\mathrm{RF}$}
For the given point $\left(\mathbf{V}^{(j)}_\mathrm{RF},\mathbf{V}^{(j)}_\mathrm{BB}\right)$, we seek to optimize $\mathbf{V}^{(j+1)}_\mathrm{BB}$ that satisfies the following condition
\begin{equation}
     g^{(j)}\left(\mathbf{V}^{(j)}_{\rm RF},\mathbf{V}^{(j+1)}_\mathrm{BB}\right) >  g^{(j)}\left(\mathbf{V}^{(j)}_{\rm RF},\mathbf{V}^{(j)}_\mathrm{BB}\right),
\end{equation}
by considering the following sub-problem for the BB TPC
\begin{subequations}\label{GM_OP:5}
\begin{align} 
&\mathop {\max }\limits_{\mathbf{V}_\mathrm{BB}} \quad    \sum_{k=1}^K {u}^{(j)}_k \ln{\left(1+\gamma^{(j)}_k\left(\mathbf{V}_\mathrm{BB}\right)\right)} \label{GM_OF:5}\\
&\mathrm{s.\hspace{1mm} t.} \quad \rho_\mathrm{th} = \mu_\mathrm{th}\left(\sum_{k=1}^{K} \left| \mathbf{a}^{H}_\mathrm{BS}({\theta})\mathbf{V}^{(j)}_{\rm RF}\mathbf{v}_{\mathrm{BB}, k} \right|^2\right)^2 \geq \widetilde{\eta}_\mathrm{th},\label{cons:PD_WSR_UP} \\ 
&\qquad \quad \|\mathbf{V}^{(j)}_\mathrm{RF}\mathbf{V}_\mathrm{BB}\|_F^2\leq P_\mathrm{t},\label{cons:TP_WSR_UP}
\end{align}
\end{subequations}
where $\widetilde{\eta}_\mathrm{th}$ is the solution obtained for the corresponding value of $\rho_\mathrm{th}$ by solving $1 - \mathcal {F}_{\chi _{2}^{2}(\rho_\mathrm{th})} \Big (\mathcal {F}_{\chi _{2}^{2}}^{-1}(1 - P_{\text {FA}}) \Big) = P_\mathrm{th}$ and $\gamma^{(j)}_k\left(\mathbf{V}_\mathrm{BB}\right) = \frac{\left\vert{\mathbf{\widetilde{h}}}^{(j)}_k\mathbf{v}_{\mathrm{BB},k}\right\vert^2}
{\sum\limits_{i\neq k}^{K}{\left\vert{\mathbf{\widetilde{h}}}^{(j)}_k \mathbf{v}_{\mathrm{ BB},i}\right\vert^2} + \sigma_{k}^{2}}$ with ${\mathbf{\widetilde{h}}}^{(j)}_k = \mathbf{h}^{H}_k\mathbf{V}^{(j)}_\mathrm{RF}$. Let us define $\bar{u} = {\mathbf{\widetilde{h}}}^{(j)}_k\mathbf{v}_{\mathrm{BB}, k}^{(j)}$ and $\bar{v} = {\sum\limits_{i\neq k}^{K}{\left\vert{\mathbf{\widetilde{h}}}^{(j)}_k \mathbf{v}_{\mathrm{BB}, i}^{(j)}\right\vert^2} + \sigma_{k}^{2}}$. 
Then, a quadratic minorizing function for objective (\ref{GM_OF:5}) at point $\mathbf{V}^{(j)}_\mathrm{BB}$ is constructed as follows
\begin{equation}\label{Minorant_1}
\begin{aligned}
\sum_{k = 1}^{K} {u}^{(j)}_k \Big(2  \Re \{a^{(j)}_{k}\mathbf{\widetilde{h}}^{(j)}_k \mathbf{v}_{\mathrm{BB}, k}\} - b^{(j)}_{k} \sum _{i = 1}^{K}{\left\vert{\mathbf{\widetilde{h}}}^{(j)}_k \mathbf{v}_{\mathrm{BB}, i}\right\vert^2} + c^{(j)}_{k}\Big),
\end{aligned}
\end{equation}  
where $a^{(j)}_{k} = \frac{\gamma^{(j)}_k\left(\mathbf{V}^{(j)}_\mathrm{BB}\right)}{{\mathbf{\widetilde{h}}}^{(j)}_k\mathbf{v}^{(j)}_{\mathrm{BB}, k}}$, $b^{(j)}_{k} = \frac{\gamma^{(j)}_k\left(\mathbf{V}^{(j)}_\mathrm{BB}\right)}{\sum_{i = 1}^{K}{\left\vert{\mathbf{\widetilde{h}}}^{(j)}_k \mathbf{v}_{\mathrm{BB}, i}\right\vert^2} +\sigma_{k}^{2}}$ and $c^{(j)}_{k} = \ln\left(1 + \gamma^{(j)}_k\left(\mathbf{V}^{(j)}_\mathrm{BB}\right)\right) -  \gamma^{(j)}_k\left(\mathbf{V}^{(j)}_\mathrm{BB}\right) - b^{(j)}_{k}\sigma_{k}^{2}$. 
Furthermore, by rearranging the terms of (\ref{Minorant_1}) and omitting the constant terms, followed by the substitution of the non-convex constraint (\ref{cons:PD_WSR_UP}) with its associated minorant (\ref{eqn:SFD}) as detailed in Section \ref{PD-BB}, the equivalent modified problem of (\ref{GM_OP:5})
is given by
\begin{subequations}\label{GM_OP:6}
\begin{align} 
\mathop {\min }\limits_{\mathbf{V}_\mathrm{BB}} &\quad \sum_{k=1}^{K}\mathbf{v}_{\mathrm{BB}, k}^{H}\Phi^{(j)} \mathbf{v}_{\mathrm{BB}, k} -2\sum_{k=1}^{K}\Re\{\mathrm{d}^{(j)}_{k}\mathbf{v}_{\mathrm{BB}, k}\} \label{GM_OF:6}\\
\mathrm{s.\hspace{1mm} t.} &\quad 2\sum_{k=1}^{K}\Re\{\mathrm{{v}}_{\mathrm{BB},k}^{(j)H}\Omega^{(j)}_{d}\mathbf{v}_{\mathrm{BB},k}\} \geq \widetilde{\omega}_\mathrm{th}^{(j)},\\ 
&\quad \sum_{k=1}^{K}\mathbf{v}_{\mathrm{BB}, k}^{H}\Omega^{(j)}_{p} \mathbf{v}_{\mathrm{BB}, k}\leq P_\mathrm{t},
\end{align}
\end{subequations}
where $\Phi^{(j)}=\sum_{i=1}^{K}{u}^{(j)}_ib^{(j)}_{i}\mathbf{\widetilde{h}}^{(j){H}}_i\mathbf{\widetilde{h}}^{(j)}_i$, $\mathrm{d}^{(j)}_{k} = {u}^{(j)}_k a^{(j)}_{k}\mathbf{\widetilde{h}}^{(j)}_k, \forall k$, $\Omega^{(j)}_{p} =\mathbf{V}^{(j){H}}_{\rm RF}\mathbf{V}^{(j)}_{\rm RF}$, $\Omega^{(j)}_{d} = \mathbf{V}^{(j){H}}_{\rm RF}{\mathbf{A}}(\theta)\mathbf{V}^{(j)}_{\rm RF}$ and $\widetilde{\omega}_\mathrm{th}^{(j)} = \omega_\mathrm{th} + \sum_{k=1}^{K}\mathrm{{v}}_{\mathrm{BB},k}^{(j)H}\omega^{(j)}\mathrm{{v}}_{\mathrm{BB},k}^{(j)}$ with $\omega =  \sqrt{\frac{\widetilde{\eta}_\mathrm{th}}{\mu_\mathrm{th}}}$. 
Since both the objective function and the constraints of the problem (\ref{GM_OP:6}) are convex, one can solve it efficiently via a standard convex solver, such as \cite{cvx}. 
\subsection{Optimization of $\mathbf{V}_\mathrm{RF}$ design for a fixed $\mathbf{V}_\mathrm{BB}$}
Next, for a given $\left(\mathbf{V}^{(j)}_\mathrm{RF}, \mathbf{V}^{(j+1)}_\mathrm{BB}\right)$, we further seek to optimize $\mathbf{V}^{(j+1)}_\mathrm{RF}$ that satisfies the following condition
\begin{equation}
     g^{(j)}\left(\mathbf{V}^{(j+1)}_\mathrm{RF},\mathbf{V}^{(j+1)}_\mathrm{BB}\right) >  g^{(j)}\left(\mathbf{V}^{(j)}_\mathrm{RF}, \mathbf{V}^{(j+1)}_\mathrm{BB}\right),
\end{equation}
by considering the following sub-problem for the RF TPC:
\begin{subequations}\label{GM_OP:7}
\begin{align} 
&\mathop {\max }\limits_{\mathbf{V}_\mathrm{RF}} \quad   \sum_{k=1}^K {u}^{(j)}_k \ln{\left(1+\gamma^{(j)}_k(\mathbf{V}_{\rm RF})\right)} \label{GM_OF:7}\\
&\mathrm{s.\hspace{1mm} t.} \quad \mu_\mathrm{th}\left(\sum_{k=1}^{K} \left| \mathbf{a}^{H}_\mathrm{BS}({\theta})\mathbf{V}_{\rm RF}\mathbf{v}^{(j+1)}_{\mathrm{BB}, k} \right|^2\right)^2 \geq \tilde{\eta}_\mathrm{th} ,\label{cons:PD_WSR_RF}\\ 
&\qquad \quad \|\mathbf{V}_\mathrm{RF}\mathbf{V}^{(j+1)}_\mathrm{BB}\|_F^2\leq P_\mathrm{t},\label{cons:TP_WSR_RF}\\
&\qquad \quad \left\vert\mathbf{V}_\mathrm{RF}(i,j)\right\vert = 1, \forall i, j,
\end{align}
\end{subequations}
where $\gamma^{(j)}_k\left(\mathbf{V}_\mathrm{RF}\right) = \mathbf{\gamma}_k\left(\mathbf{V}_\mathrm{RF}, \mathbf{V}^{(j+1)}_\mathrm{BB}\right)$. Furthermore, by using the transformation $\mathbf{h}_k^H \mathbf{V}_\mathrm{ RF}\mathbf{v}^{(j+1)}_{\mathrm{BB}, i} = \left[\left(\mathbf{v}^{(j+1)}_{\mathrm{BB},i}\right)^{T} \otimes \mathbf{h}_k^H\right]\text{vec}\ (\mathbf{V}_{\text{RF}})$ and by definition of $\boldsymbol{\phi}$ as $\boldsymbol{\phi} = \mathrm{vec}\left(\mathbf{V}_\mathrm{RF}\right)$, we rewrite $\gamma^{(j)}_k(\mathbf{V}_\mathrm{RF})$ in terms of $\boldsymbol{\phi}$ as $\gamma^{(j)}_k\left(\boldsymbol{\phi}\right) = \frac{\left\vert{\mathbf{\widetilde{h}}}^{(j+1)}_{k,k}\boldsymbol{\phi}\right\vert^2}
{\sum\limits_{i\neq k}^{K}{\left\vert{\mathbf{\widetilde{h}}}^{(j+1)}_{k,i}\boldsymbol{\phi}\right\vert^2} + \sigma_{k}^{2}}$, where ${\mathbf{\widetilde{h}}}^{(j+1)}_{k,i} \triangleq \left[(\mathbf{v}^{(j+1)}_{\mathrm{BB},i})\right]^T\otimes \mathbf{h}_k^H$.
Given this transformation, the quadratic
minorizing function of $\gamma^{(j)}_k(\boldsymbol{\phi})$ at point $\boldsymbol{\phi}^{(j)} = \text{vec}(\mathbf{V}^{(j)}_\mathrm{RF})$ is obtained by considering $ \bar{u} = {\mathbf{\widetilde{h}}}^{(j+1)}_{k,k}\boldsymbol{\phi}^{(j)}$ and $\bar{v} = {\sum\limits_{i\neq k}^{K}{\left\vert{\mathbf{\widetilde{h}}}^{(j+1)}_{k,i}\boldsymbol{\phi}^{(j)}\right\vert^2} + \sigma_{k}^{2}}$ as follows
\begin{equation}\label{Minorant_2}
\begin{aligned}
\sum _{k = 1}^{K} {u}^{(j)}_k \left(2  \Re \{\tilde{a}^{(j)}_{k}\mathbf{\widetilde{h}}^{(j+1)}_{k,k} \boldsymbol{\phi}\} - \tilde{b}^{(j)}_{k} \sum _{i = 1}^{K}{\left\vert{\mathbf{\widetilde{h}}}^{(j+1)}_{k,i} \boldsymbol{\phi}\right\vert^2} + \tilde{c}^{(j)}_{k}\right),
\end{aligned}
\end{equation}  
where $\tilde{a}^{(j)}_{k} = \frac{\gamma^{(j)}_k(\boldsymbol{\phi}^{(j)})}{{\mathbf{\widetilde{h}}}^{(j+1)}_{k,k}\boldsymbol{\phi}^{(j)}}$, $\tilde{b}^{(j)}_{k} = \frac{\gamma^{(j)}_k(\boldsymbol{\phi}^{(j)})}{\sum _{i = 1}^{K}{\left\vert{\mathbf{\widetilde{h}}}^{(j+1)}_{k,i} \boldsymbol{\phi}^{(j)}\right\vert^2} +\sigma_{k}^{2}}$ and $\tilde{c}^{(j)}_{k} = \ln\left(1 + \gamma^{(j)}_k\left(\boldsymbol{\phi}^{(j)}\right)\right) -  \gamma^{(j)}_k\left(\boldsymbol{\phi}^{(j)}\right) - \tilde{b}^{(j)}_{k}\sigma_{k}^2$. To simplify it further, we rewrite (\ref{Minorant_2}) in a compact form as follows
\begin{equation}\label{Minorant_3}
\begin{aligned}
 2  \Re \{\mathbf{p}^{(j)}\boldsymbol{\phi}\}  - \boldsymbol{\phi}^H \mathbf{E}^{(j)}\boldsymbol{\phi} + {c}^{(j)},
\end{aligned}
\end{equation} 
where $\mathbf{E}^{(j)} = \sum _{k = 1}^{K} {u}^{(j)}_k \tilde{b}^{(j)}_{k} \left(\sum _{i = 1}^{K}(\mathbf{\widetilde{h}}^{(j+1)}_{k,i})^{H}\mathbf{\widetilde{h}}^{(j+1)}_{k,i}\right)$,  $\mathbf{{p}}^{(j)} = \sum _{k = 1}^{K} {u}^{(j)}_k \tilde{a}^{(j)}_{k}\mathbf{\widetilde{h}}^{(j+1)}_{k,k}$, and ${c}^{(j)} = \sum _{k = 1}^{K}{u}^{(j)}_k\tilde{c}^{(j)}_{k}$.
Next, we apply similar transformations to constraints (\ref{cons:PD_WSR_RF}) and (\ref{cons:TP_WSR_RF}) as illustrated in Section \ref{PD-RF}, which reformulates problem (\ref{GM_OP:7}) as follows
\begin{subequations}\label{GM_OP:8}
\begin{align}
\min \limits _{\boldsymbol{\phi}} \quad & \boldsymbol{\phi}^H \mathbf{E}^{(j)} \boldsymbol{\phi} - 2  \Re \{\mathbf{p}^{(j)}\boldsymbol{\phi}\} \\ 
\mathrm{s.~t.}\ & \boldsymbol{\phi}^H \boldsymbol{\Lambda}^{(j)}_{d} \boldsymbol{\phi} \geq \omega_\mathrm{th}, \label{cons:jite_6}\\ 
& \boldsymbol{\phi}^H \boldsymbol{\Pi}^{(j)}_{p} \boldsymbol{\phi} \leq P_\mathrm{t}, \label{cons:jite_7}\\ 
& \vert\boldsymbol{\phi}(n)\vert=1, \forall n,
\end{align}
\end{subequations}
where $\boldsymbol{\Lambda}^{(j)}_{d} = \sum \limits_{\begin{subarray}{l} i = 1 \end{subarray}}^{K} \left[\left( \mathbf{v}^{(j+1)}_{\mathrm{BB},i} (\mathbf{v}^{(j+1)}_{\mathrm{BB},i})^{H} \right)^{T} \otimes {\mathbf{A}}(\theta)\right]$ and $\boldsymbol{\Pi}^{(j)}_{p} = \left[ \mathbf{\mathbf{V}^{(j+1)}_\mathrm{BB}}\left(\mathbf{V}^{(j+1)}_\mathrm{BB}\right)^H\right]^{T} \otimes \mathbf{I}_{N_{t}}$. 
Subsequently, we adopt the PRCG algorithm, which converts the above constrained problem (\ref{GM_OP:8}) into an unconstrained problem on the Riemannian manifold upon adding the constraints (\ref{cons:jite_6}) and (\ref{cons:jite_7}) into the objective function. Thus, the equivalent penalized problem is given by
\begin{equation}\label{RCG_2}
\begin{aligned}
& \min_{\boldsymbol{\phi}}~ \mathcal{F}^{(j)}(\boldsymbol{\phi}) \\
&= \boldsymbol{\phi}^H \mathbf{E}^{(j)} \boldsymbol{\phi} - 2  \Re \{\mathbf{p}^{(j)}\boldsymbol{\phi}\} + \lambda \Big( \nu^{(j)}_d\left(\boldsymbol{\phi}\right)+\chi^{(j)}_p\left(\boldsymbol{\phi}\right)\Big)\\ 
&\text{s.~t.} \quad \vert\boldsymbol{\phi}(n)\vert=1, \forall n,
\end{aligned}
\end{equation}
where $\nu^{(j)}_d\left(\boldsymbol{\phi}\right) \triangleq \left(\max\{0,\omega_\mathrm{th} - \boldsymbol{\phi}^H \boldsymbol{\Lambda}^{(j)}_{d} \boldsymbol{\phi}\}\right)^2$, $\chi^{(j)}_p\left(\boldsymbol{\phi}\right) \triangleq \left(\max\{0, \boldsymbol{\phi}^H \boldsymbol{\Pi}^{(j)}_{p} \boldsymbol{\phi} - P_\mathrm{t} \}\right)^2$ and $\lambda$ is the penalty factor. 
Next, to solve the above problem (\ref{RCG_2}) on the Riemannian manifold for a given $\lambda$, the Euclidean gradient of $\mathcal{F}^{(j)}(\boldsymbol{\phi})$ is given by
\begin{equation}  
\nabla \mathcal{F}^{(j)}(\boldsymbol{\phi})  = 2 \mathbf{E}^{(j)} \boldsymbol{\phi} - 2(\mathbf{p}^{(j)})^{H} + \lambda \left( \boldsymbol{\xi}^{(j)}_d + \boldsymbol{\xi}^{(j)}_{p}\right),  
\end{equation}
where the quantities $\boldsymbol{\xi}^{(j)}_d$ and $\boldsymbol{\xi}^{(j)}_p$ are defined as
\begin{subequations}\label{eqn:EG_SR} 
\begin{align}
&\boldsymbol{\xi}^{(j)}_d = \begin{cases}
     4\left(\boldsymbol{\phi}^H \boldsymbol{\Lambda}^{(j)}_{d} \boldsymbol{\phi}-\omega_\mathrm{th}\right)\boldsymbol{\Lambda}^{(j)}_{d} \boldsymbol{\phi},&\omega_\mathrm{th} \geq \boldsymbol{\phi}^H \boldsymbol{\Lambda}^{(j)}_{d} \boldsymbol{\phi}, \\
    0, & \omega_\mathrm{th} < \boldsymbol{\phi}^H \boldsymbol{\Lambda}^{(j)}_{d} \boldsymbol{\phi},
\end{cases}\\
&\boldsymbol{\xi}^{(j)}_p = \begin{cases}
     4\left( \boldsymbol{\phi}^H \boldsymbol{\Pi}^{(j)}_{p} \boldsymbol{\phi} - P_\mathrm{t}\right)\boldsymbol{\Pi}^{(j)}_{p} \boldsymbol{\phi}, & \boldsymbol{\phi}^H \boldsymbol{\Pi}^{(j)}_{p} \boldsymbol{\phi}  \geq P_\mathrm{t}, \\ 0, & \boldsymbol{\phi}^H \boldsymbol{\Pi}_{p}^{(j)} \boldsymbol{\phi}  < P_\mathrm{t}.
\end{cases}
\end{align}
\end{subequations}
Thus, the optimal problem solution of (\ref{GM_OP:7}) is achieved via the PRCG algorithm, where the penalty parameter $\lambda$ progressively increases until the constraints are met. 
Hence, we alternatively optimize the BB TPC $\mathbf{V}_\mathrm{BB}$ and RF TPC $\mathbf{V}_\mathrm{RF}$ using the proposed MM-Alt method. Furthermore, Algorithm 4 presents a pseudo-code of the proposed MM-Alt method for jointly optimizing the RF and BB TPCs for solving the GMR-max problem. 

Observe that Algorithm 4 employs Algorithm 1 and Algorithm 2 for optimizing the BB and RF TPCs, respectively, in an alternating fashion. Therefore, the overall computational complexity of Algorithm 4 is given by $\mathcal{O}\left(\mathcal{I}_\mathrm{o}\left(N^{3.5}_{RF}K^{3.5} + \mathcal{I}_\mathrm{r}N^2_{t}N^2_{RF}\frac{1}{\epsilon^2_1}\right)\right)$,
where $\mathcal{I}_\mathrm{o}$ is the number of iterations required for the convergence of Algorithm 4.

Note that the proposed designs are also applicable to the partially-connected hybrid MIMO architecture, where the RF TPC matrix adopts a block-diagonal form \cite{mmWave_ISAC_2}, affecting only the RF TPC design. Since both optimization problems (\ref{PD_OP:1})
and (\ref{GMR_OP:1}) use the proposed PRCG algorithm, the block-diagonal RF TPC can be vectorized into the tractable forms of (\ref{PD_OP:9}) and (\ref{GM_OP:8}), allowing the PRCG method to efficiently solve the problems while preserving the architecture's structural constraints.

\begin{algorithm}[t]
\caption{MM-Alt method for solving $\mathcal{P}_2$}
 \textbf{Input:} $P_\mathrm{t}, P_\mathrm{FA}$, $P_\mathrm{th}$
\begin{algorithmic}[1]
\label{alg:MM_BCD}
\State \textbf{initialize:} $j = 0 $, feasible TPCs $\mathbf{V}^{(j)}_\mathrm{RF}$ and $ \mathbf{V}^{(j)}_\mathrm{BB}$
         \Repeat   
                \State compute ${u}^{(j)}_k ,\ \forall k$ using (\ref{eqn:AW})
                \State obtain $\mathbf{V}^{(j + 1)}_\mathrm{BB}$ for given $\mathbf{V}^{(j )}_\mathrm{RF}$ by solving (\ref{GM_OP:6}).
                \State find $\mathbf{V}^{(j + 1)}_\mathrm{RF}$ for given $\mathbf{V}^{(j + 1)}_\mathrm{BB}$ by solving (\ref{RCG_2}) via Algorithm 2.
           \Until $f_\mathrm{GM}(\mathbf{V}_{\rm RF},\mathbf{V}_\mathrm{BB})$ converges.
\end{algorithmic}
\end{algorithm}

\section{\uppercase{Simulation Results}}\label{simulation results}
In this section, we present our simulation results to evaluate both the detection and communication performance for demonstrating the effectiveness of the proposed algorithms for HBF design in an ISAC-enabled mmWave system. Throughout the simulations, we use the following settings, unless stated otherwise. 
The ISAC BS is configured with a ULA having $N_\mathrm{t} = 128$ antennas and $N_\mathrm{RF} \in \{8, 16\}$ RFCs. We consider $K = 4$ CUs, which are positioned at distances of $40$ m, $30$ m, $20$ m, and $10$ m in a circular area of radius $50$ m from the ISAC BS, at angles of $-60^\circ, -30^\circ, 30^\circ$, and $60^\circ$, respectively. The RT is located at $0^\circ$. 
Furthermore, the mmWave channel gain is modeled as $\beta_{\ell,k} \sim \mathcal{CN}(0, 10^{-0.1PL(d_m)})$, where $PL(d_k)$ represents the path loss gain and it is given by \cite{HBF_8}
\begin{equation}\label{eqn:path loss model}
\begin{aligned}
PL(d_k)\hspace{0.02in}[\rm dB] = \varepsilon + 10\varphi\log_{10}(d_k)+\varpi,
\end{aligned}
\end{equation}
where $d_k$ is the associated distance of the $k$th CU from the ISAC BS, and the quantities $\varepsilon$, $\varphi$ and $\varpi$ are given by $\varepsilon=61.4$, $\varphi=2$ and $\varpi \in {\cal CN}(0,\sigma_{\rm \varpi}^2)$, with $\sigma_{\rm \varpi}=5.8 \hspace{0.02 in}{\rm dB}$ \cite{HBF_8}.
Moreover, the system operates at the carrier frequency of $28$ GHz with a bandwidth of $251.1886$ MHz and a total power budget of $P_\mathrm{t}$ = 30dBm. Thus, the noise variance at each CU is set as $ \sigma_{k}^{2} = -174+ 10 \log_{10}B = -90$ dBm. The radar cross-section is generated as $\alpha \in \mathcal{CN}(0, \sigma^2_{\alpha})$ with $\sigma^2_{\alpha} = -90$ dB and the noise variance of the echo signal is set as $\sigma_{k}^{2}$ = -60 dBm. 
While we model $\alpha$ as $\alpha \in \mathcal{CN}(0, \sigma^2_{\alpha})$, incorporating distance-dependent path loss is an interesting direction for future work.
\subsection{Trade-off between sensing and communication via PD-max}
In this subsection, we investigate the performance analysis of the proposed Algorithms toward designing the HBF via PD-max optimization. We consider the minimum SINR requirement for each CU to be the same as $\Gamma_k = \Gamma, \forall k$, and set the parameters $\Gamma = 15$ dB, $P_\mathrm{FA}= 10^{-6}$ and $\eta = 0.975$, unless stated otherwise. Furthermore, we compare the proposed Bi-Alt scheme to the following benchmarks:
\begin{itemize}
    \item {\bf Scheme 1 (Sensing-only)}: For this scheme, we allocate the total available power exclusively to sensing. Thus, the BB and RF TPCs are optimized via solving problem $\mathcal{P}_1$ by removing the SINR constraint (\ref{cons:SINR}).
    \item {\bf Scheme 2 (FDB)}: For this scheme, we employ the FDB scheme to solve $\mathcal{P}_1$, which requires $N_\mathrm{RF} = N_\mathrm{t}$.
    \item {\bf Scheme 3 (HBF, two-stage)}: For this scheme, we employ the two-stage design \cite{mmISAC_1}, where we first obtain the FDB corresponding to $\mathcal{P}_1$ and subsequently optimize the BB and RF TPCs via minimizing the Euclidean distance between the FDB and HBF beamfomers.
    \item {\bf Scheme 4 (HBF, OMP)}: For this scheme, we employ the OMP algorithm \cite{mm_OMP} to optimize the BB and RF TPCs in the inner layer of the Bi-Alt algorithm.
\end{itemize}
\subsubsection{Convergence behavior}
\begin{figure}[t]
\setkeys{Gin}{width=\linewidth}
    \begin{subfigure}[t]{0.24\textwidth}
    \includegraphics[width=1.1\textwidth]{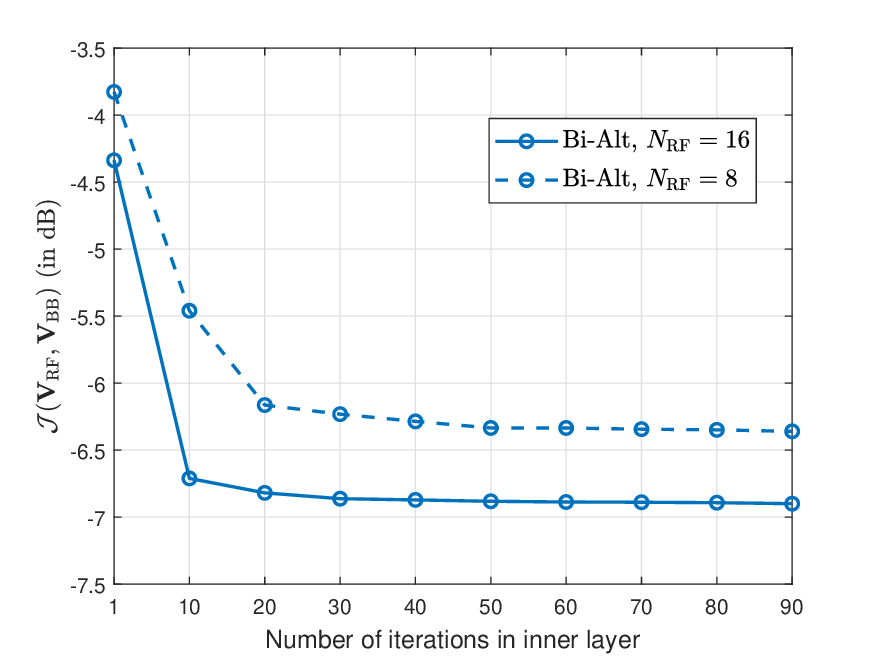}
    \caption{} \label{fig:R1_1}
\end{subfigure}
\begin{subfigure}[t]{0.24\textwidth}
    \includegraphics[width=1.1\textwidth]{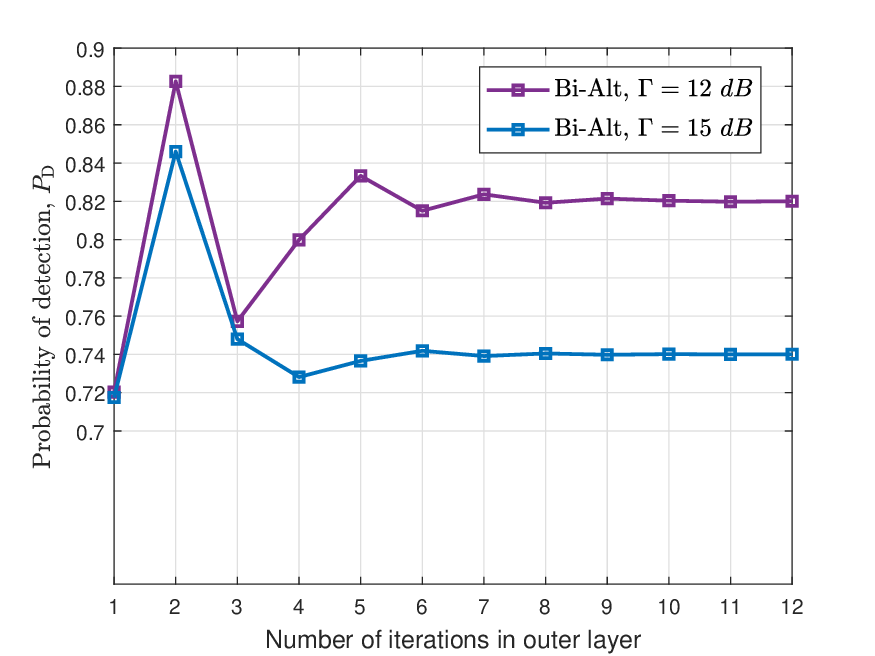}
    \caption{} \label{fig:R3_4}
\end{subfigure}
\caption{Convergence behavior of the proposed Bi-Alt algorithm in (a) inner layer; (b) outer layer.}
\vspace{-0.7cm}
\end{figure}

Here we characterize the convergence behavior of the proposed Bi-Alt Algorithm 3 for solving (\ref{PD_OP:1}), particularly for larger antenna arrays and increased RF chains. The inner loop of the algorithm follows a BCD approach, iteratively updating the RF and BB TPCs, $\mathbf{F}_\mathrm{RF}$ and $\mathbf{F}_\mathrm{BB}$, to minimize the objective function $\mathcal{J}\left(\mathbf{V}_\mathrm{RF}, \mathbf{V}_\mathrm{BB}\right)$. At the $(j+1)$th iteration, the monotonic convergence property is ensured as follows:
\begin{equation} 
\begin{aligned}
&\mathcal{J}(\mathbf{F}^{(j + 1)}_\mathrm{RF}, \mathbf{F}^{(j + 1)}_\mathrm{BB})  \leq\mathcal{J}(\mathbf{F}^{(j)}_\mathrm{RF}, \mathbf{F}^{(j + 1)}_\mathrm{BB}) \leq \mathcal{J}(\mathbf{F}^{(j)}_\mathrm{RF}, \mathbf{F}^{(j)}_\mathrm{BB}),
\end{aligned}
\end{equation}
where the RF TPC $\mathbf{F}^{(j)}_\mathrm{RF}$ is optimized via the PRCG method. The PRCG method iteratively refines the feasibility points while ensuring a non-increasing sequence of objective function values. Given the outer loop structure, the slack variable $\eta$ is updated via the bisection search method until convergence is achieved. 
Furthermore, Fig. \ref{fig:R1_1} shows the convergence behavior of the inner layer of the proposed Bi-Alt algorithm with $\epsilon_3 = 10^{-3}$, and $N_\mathrm{RF}= \{8,~16\}$, which minimizes the function $\mathcal{J}\left(\mathbf{V}_\mathrm{RF}, \mathbf{V}_\mathrm{BB}\right)$. 
As shown in the figure, the function $\mathcal{J}\left(\mathbf{V}_\mathrm{RF}, \mathbf{V}_\mathrm{BB}\right)$ gradually decreases and reaches its minimum value within a few iterations, which shows the convergence of the inner layer of the Bi-Alt algorithm. Observe that the function $\mathcal{J}\left(\mathbf{V}_\mathrm{RF}, \mathbf{V}_\mathrm{BB}\right)$ associated with $N_\mathrm{RF}=16$ converges faster than for $N_\mathrm{RF}=8$, as the influence of BB TPC over the RF TPCs is greater for a large number of RFCs.
Furthermore, Fig. \ref{fig:R3_4} illustrates the convergence behavior of the outer layer of the Bi-Alt with $\epsilon_4 = 10^{-6}$, which corresponds to the bisection search algorithm. As seen from the figure, the PD $P_\mathrm{D}$ saturates within $10$ iterations for both $\Gamma = 12$ dB and $15$ dB, which verifies the rapid convergence of the proposed Bi-Alt algorithm.
Note that the nonmonotonic behavior in Fig. \ref{fig:R3_4} is due to the bisection search method in the outer layer of the Bi-Alt Algorithm, which iteratively updates the slack variable while ensuring feasibility under the transmit power constraint.
\begin{figure*}[t]
\centering
\begin{subfigure}{.65\columnwidth}
\includegraphics[width=\columnwidth]{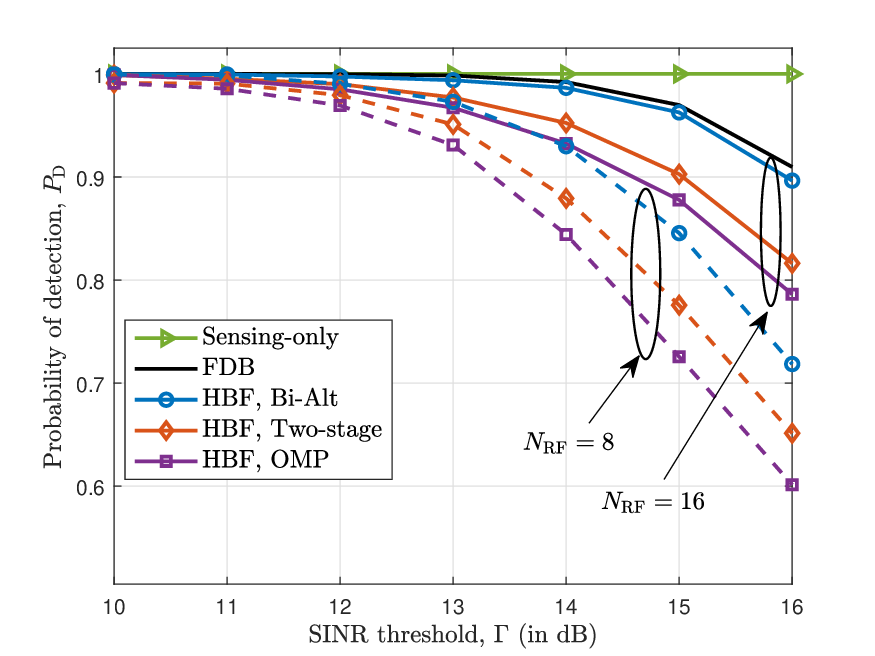}%
\caption{}
\label{fig:R2_2}
\end{subfigure}
\begin{subfigure}{.65\columnwidth}
\includegraphics[width=\columnwidth]{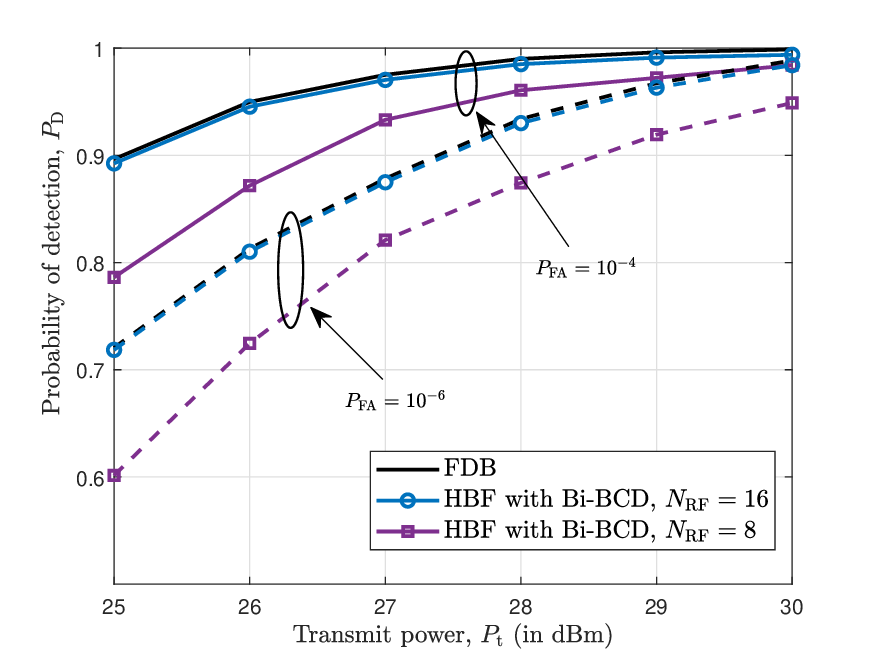}%
\caption{}
\label{fig:R3_3_1}
\end{subfigure}%
\begin{subfigure}{.65\columnwidth}
\includegraphics[width=\columnwidth]{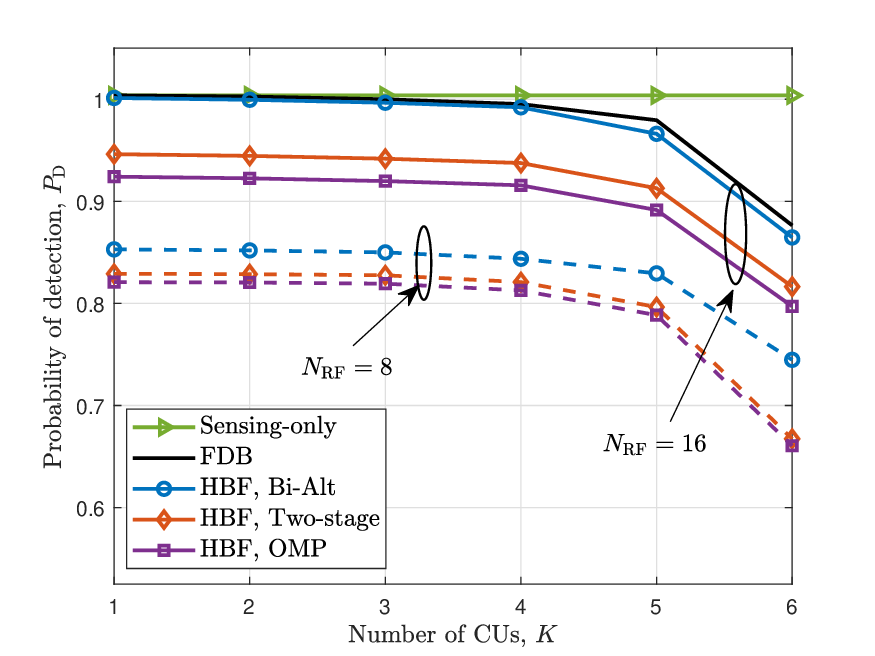}%
\caption{}
\label{fig:R4_4}
\end{subfigure}%
\caption{Probability of detection versus (a) SINR threshold; 
(b) transmit power; (c) number of CUs.}
\vspace{-5mm}
\end{figure*}
\subsubsection{Probability of detection versus SINR threshold}
Fig. \ref{fig:R2_2} investigates the behavior of the PD with respect to the
SINR threshold for various values of the number of RFCs. 
As shown in the figure, the PD decreases upon increasing the SINR threshold, which reveals the trade-off between the sensing and communication tasks. This trade-off is intuitive, since more power is radiated toward the CUs to meet the higher SINR requirements, which reduces the power available for RT detection.
Observe that the PD of the sensing-only scheme is independent of the SINR threshold and acts as the upper bound for the sensing performance.
Furthermore, the proposed Bi-Alt scheme having $N_\mathrm{RF}=16$ performs close to the FDB scheme and outperforms the two-stage HBF and OMP benchmarks for both $N_\mathrm{RF}=8$ and $16$. This shows the efficacy of the proposed PRCG and SCA algorithms employed in our Bi-Alt scheme for the optimization of the RF and BB TPCs, respectively. 


\subsubsection{Probability of detection versus transmit power}
To further investigate the performance of the proposed HBF design based on the Bi-Alt algorithm, we plot the probability of detection versus transmit power in Fig. \ref{fig:R3_3_1}.
As shown in the figure, the probability of detection increases monotonically upon increasing the transmit power, which is a benefit of having a higher power for RT detection for a fixed SINR requirement. Observe that the probability of detection for the proposed scheme associated with $N_\mathrm{RF}=16$ RFCs approaches that of the FDB scheme, and it is enhanced in comparison to the benchmarks for both $N_\mathrm{RF}=8$ and $16$ at high as well as low transmit powers. Interestingly, the performance gap between the proposed scheme having $N_\mathrm{RF} = 8$ and the FDB scheme reduces upon increasing the transmit power. Hence, one can further reduce the number of RFCs in mmWave ISAC systems for achieving the desired probability of detection by increasing the transmit power.


\subsubsection{Probability of detection versus the number of CUs}
In Fig. \ref{fig:R4_4}, we investigate the impact of the number of CUs on the probability of detection for a fixed transmit power of $P_\mathrm{t} = 30~$ dBm. As shown in the figure, the performance of the system degrades upon increasing the number of CUs, which is due to the increasing SINR requirements of the additional CUs.  
Furthermore, the proposed Bi-Alt scheme performs close to the FDB for $N_\mathrm{RF}=16$, even for an increasing number of CUs in the system. Furthermore, it is superior to the other two benchmarks for both $N_\mathrm{RF}=8$ and $N_\mathrm{RF}=16$. Observe that the performance of the two-stage HBF and of OMP having $N_\mathrm{RF}=8$ degrades sharply after $K=5$ CUs due to the resultant reduced HBF gain toward the RT. By contrast, the performance of the Bi-Alt scheme with $N_\mathrm{RF}=8$ does not degrade sharply, which shows the efficacy of the SCA and PRCG algorithms. 
Moreover, it is suggested that for a fixed transmit power, one has to increase the number of RFCs upon increasing the number of CUs to achieve a higher probability of detection. 

\subsection{Trade-off between sensing and communication via GMR-max}
In this subsection, we evaluate the performance of the proposed HBF design based on the MM-Alt algorithm, which ensures rate-fairness among the CUs.
Unless stated otherwise, we set the parameters for the GMR-max scheme as $P_\mathrm{t} = 30~$dBm, $P_\mathrm{FA} = 10^{-6}$ and $P_\mathrm{th}=0.975$.
Furthermore, we compare the proposed MM-Alt scheme to the following techniques to reveal interesting insights pertaining to HBF designs:
\begin{itemize}
     \item {\bf Scheme 1 (Comm-only)}: The available transmit power is used for the CUs only. Therefore, we set $P_\mathrm{th}=0$ for optimization of the BB and RF TPCs via problem $\mathcal{P}_2$.
    \item {\bf Scheme 2 (FDB)}: This scheme employs the fully digital beamformer to solve $\mathcal{P}_2$.
    \item {\bf Scheme 3 (MMR-max)}: This scheme optimizes the BB and RF TPCs to maximize the minimum CU rate by solving the following optimization problem:
    \begin{subequations}\label{MM_OP:1}
        \begin{align} 
        &\mathop {\max }\limits_{\mathbf{V}_\mathrm{RF}, ~\mathbf{V}_\mathrm{BB}} \quad  \min_{k=1,\hdots,K} R_k(\mathbf{V}_{\rm RF},\mathbf{V}_\mathrm{BB}) \label{MM_OF:1}\\
        &\mathrm{s.\hspace{1mm} t.} \quad \text{(\ref{cons:TP}), (\ref{cons:RF}), and (\ref{cons:Pd})}.
        \end{align}
    \end{subequations} 
    We solve the problem (\ref{MM_OP:1}) seen above by transforming it into a feasible problem via the introduction of an auxiliary variable for $R_k(\mathbf{V}_{\rm RF},\mathbf{V}_\mathrm{BB})$, and, subsequently employed the Bi-Alt scheme for optimizing the BB, RF TPCs, and the auxiliary variable via the SCA, PRCG and binary search algorithms.
    \item {\bf Scheme 4 (SR-max)}:  For this scheme, we optimize the BB and RF TPCs to maximize the sum rate of the system. The corresponding pertinent optimization problem for SR-max is given by
    \begin{subequations}\label{SR_OP:3}
        \begin{align} 
        &\mathop {\max }\limits_{\mathbf{V}_\mathrm{RF}, ~\mathbf{V}_\mathrm{BB}} \quad   \sum_{k=1}^K R_k(\mathbf{V}_{\rm RF},\mathbf{V}_\mathrm{BB}) \label{SR_OF:1}\\
        &\mathrm{s.\hspace{1mm} t.} \quad \text{(\ref{cons:TP}), (\ref{cons:RF}), and (\ref{cons:Pd})}.
        \end{align}
    \end{subequations}
    The optimal solution of the SR-max framework above is solved using the proposed MM-Alt algorithm by fixing the weight of the CU as ${u}^{(j)}_k = 1,\ \forall k$.
\end{itemize}

\subsubsection{Convergence behavior}
\begin{figure}[t]
\centering
\includegraphics[width = 6.8cm]{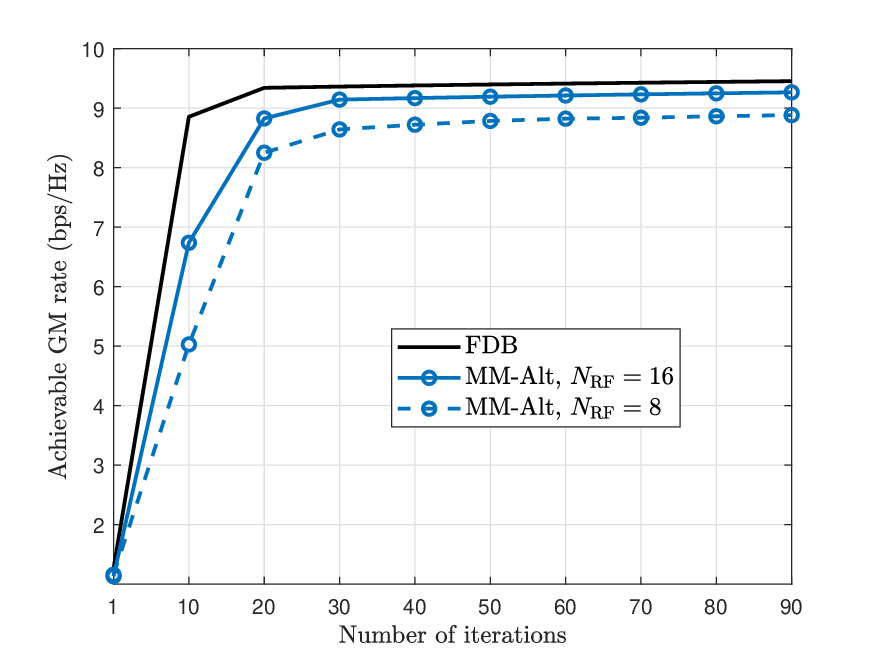}
\caption{Convergence behavior of the proposed MM-Alt algorithm.} 
\label{fig:R11}
\end{figure}
For the MM-Alt algorithm, the overall convergence behavior follows the same principles as Bi-Alt, with the key distinction that the bisection search step is omitted. Since the RF TPC significantly influences the overall convergence, and it is optimized using the PRCG method, the algorithm remains stable even for large-scale mmWave systems.
Moreover, Fig. \ref{fig:R11} shows the convergence behavior of the proposed MM-Alt algorithm conceived for maximizing the GM rate of the CUs. As shown in the figure, the GM rate of the system saturates within a few iterations for both $N_\mathrm{RF}=8$ and $16$, which verifies the convergence of the MM-Alt scheme. Furthermore, the achievable GM rate increases and achieves the rate of an ideal FDB scheme upon increasing  $N_\mathrm{RF}$, which validates the efficacy of the proposed MM-Alt scheme.

\subsubsection{Achievable GM rate versus the probability of detection threshold}

\begin{figure*}[t]
\centering
\begin{subfigure}{.65\columnwidth}
\includegraphics[width=\columnwidth]{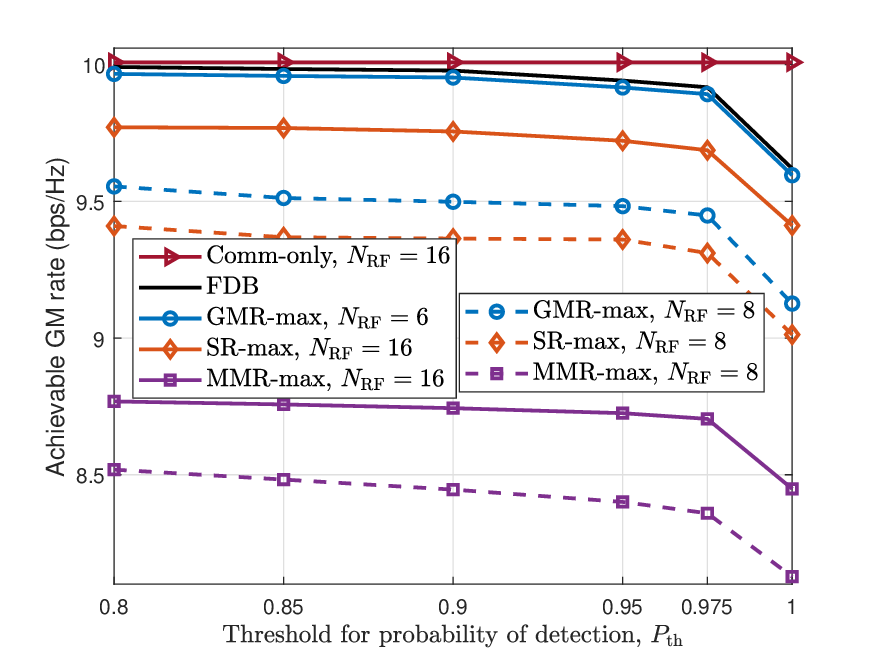}%
\caption{}
\label{fig:R5_1}
\end{subfigure}
\begin{subfigure}{.65\columnwidth}
\includegraphics[width=\columnwidth]{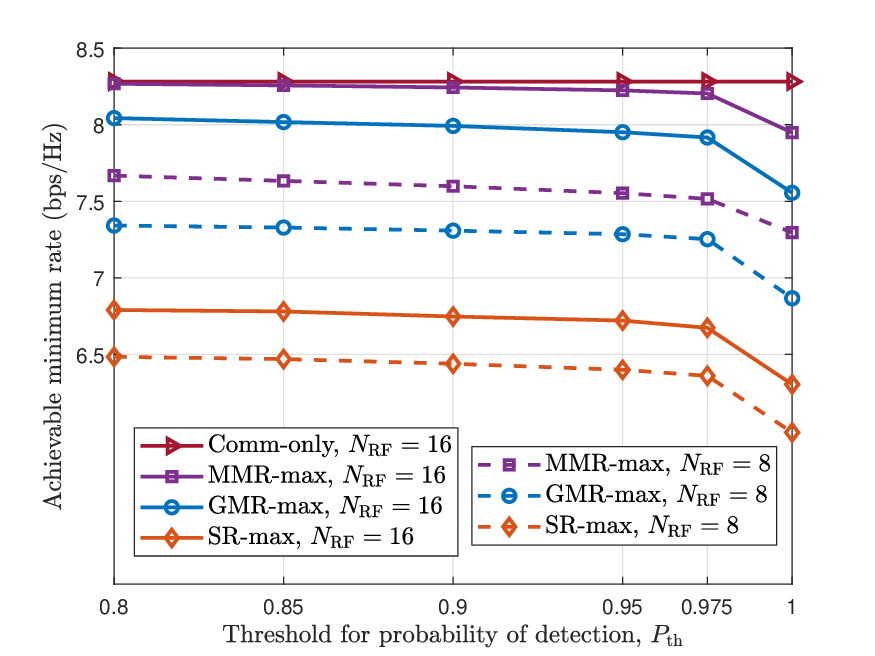}%
\caption{}
\label{fig:R6_1}
\end{subfigure}%
\begin{subfigure}{.65\columnwidth}
\includegraphics[width=\columnwidth]{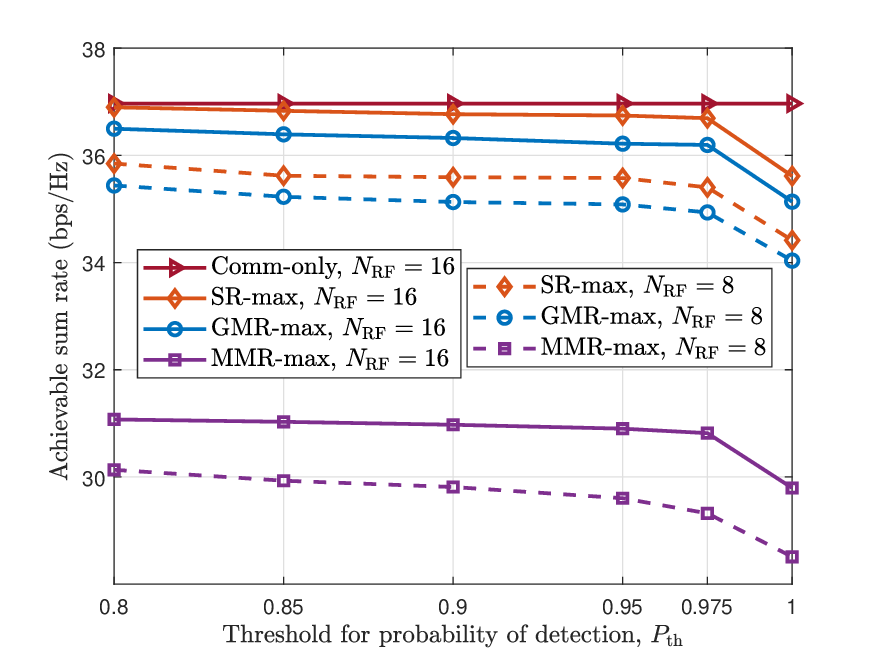}%
\caption{}
\label{fig:R7_1}
\end{subfigure}%
\caption{(a) Achievable GM rate versus probability of detection; 
(b) Achievable minimum rate versus probability of detection; (c) Achievable sum rate versus probability of detection.}
\vspace{-5mm}
\end{figure*}


\begin{figure*}[t]
\centering
\begin{subfigure}{.65\columnwidth}
\includegraphics[width=\columnwidth]{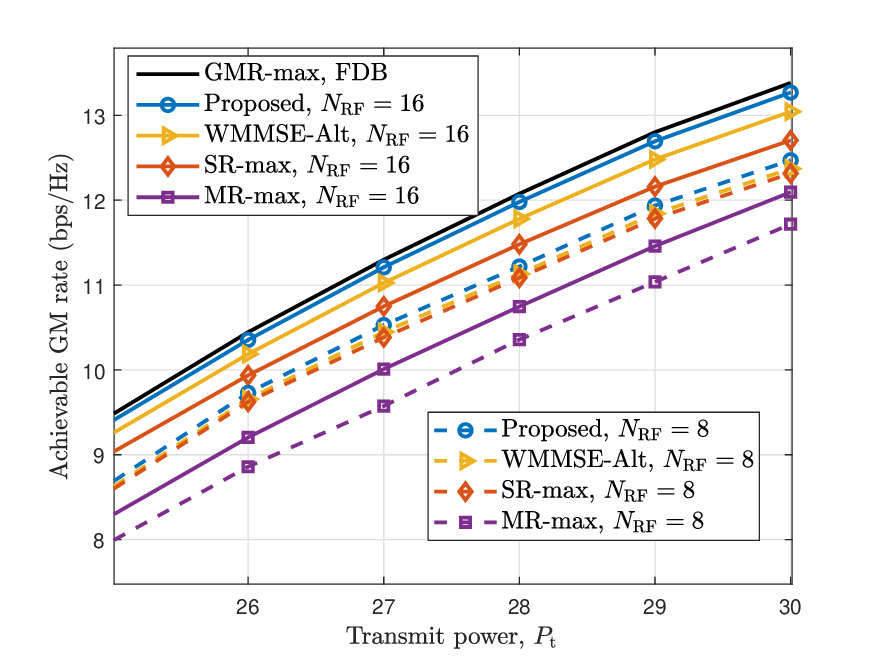}%
\caption{}
\label{fig:R8_1}
\end{subfigure}
\begin{subfigure}{.65\columnwidth}
\includegraphics[width=\columnwidth]{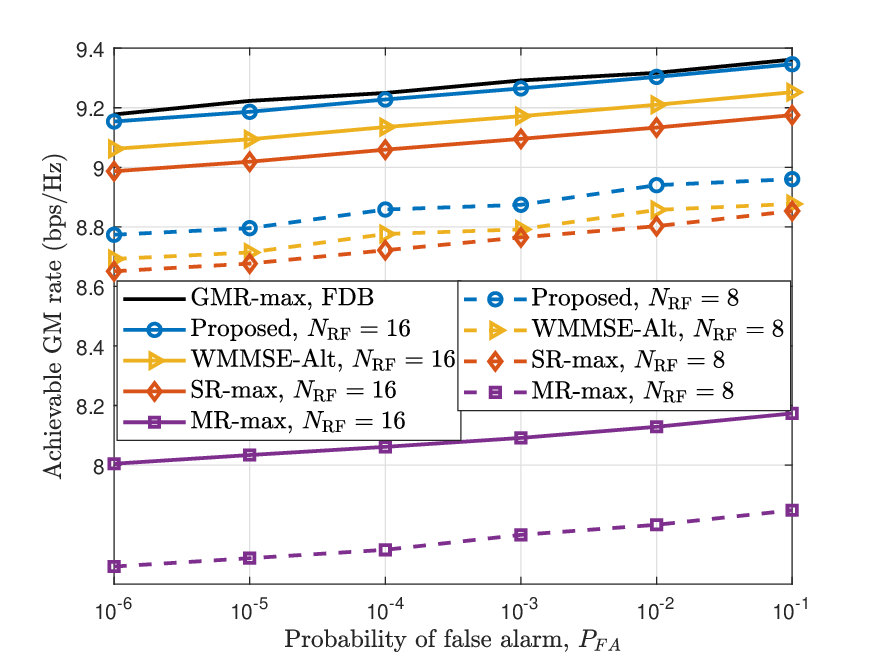}%
\caption{}
\label{fig:R9_1}
\end{subfigure}%
\begin{subfigure}{.65\columnwidth}
\includegraphics[width=\columnwidth]{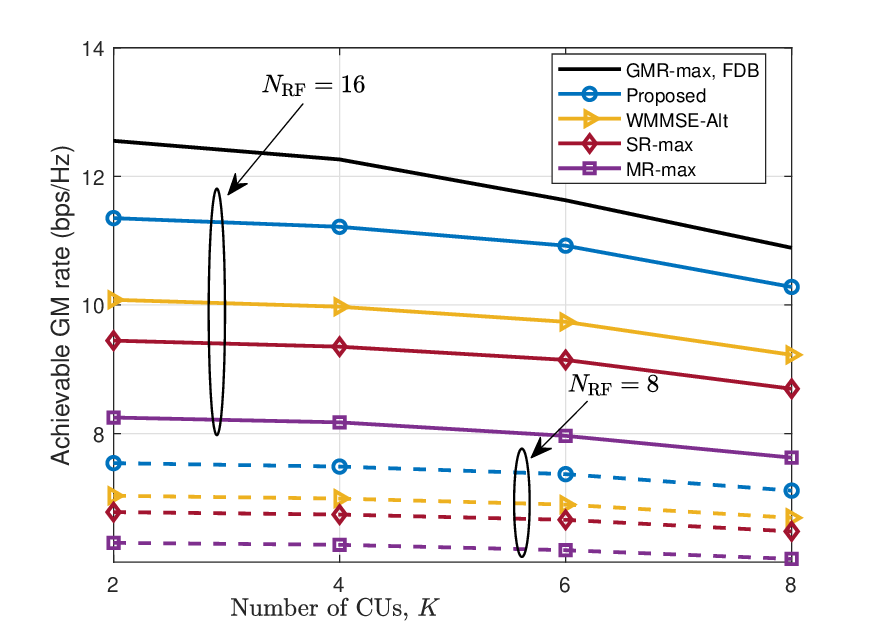}%
\caption{}
\label{fig:R15_111}
\end{subfigure}%
\caption{Achievable GM rate versus (a) transmit power; (b) probability of false alarm; (c) number of CUs.}
\vspace{-5mm}
\end{figure*}

In Fig. \ref{fig:R5_1}, we plot the achievable GM rate of the system versus the probability of detection threshold $P_\mathrm{th}$ for $N_\mathrm{RF}=\{8,~ 16\}$. The achievable GM rate of the system decreases upon increasing $P_\mathrm{th}$, due to the radiation of lower power towards the CUs.
Observed that the GM rate remains nearly constant for $P_\mathrm{th}$ between 0.8 and 0.95, as ample power is available for communication. However, for $P_\mathrm{th}>0.95$, more power is allocated for sensing, leading to a significant decline in the GM rate.
Furthermore, the GM rate of the Comm-only scheme remains unaffected by $P_\mathrm{th}$ and acts as an upper bound for the communication performance.
Interestingly, the performance of the MM-Alt algorithm with only six RFCs achieves a performance close to that of the FDB scheme, which verifies the efficacy of the SCA and PRCG algorithms, and renders it eminently suitable for the practical use cases of ISAC-enabled mmWave systems. 
Furthermore, the achievable GM rate of the GMR-max scheme is higher than that of the MMR- and SR-max schemes, which arises due to the allocation of power based on the GMR-max criteria. Therefore, the proposed MM-Alt scheme is eminently suitable for maximizing the GM rate of the system.   

\subsubsection{Achievable minimum rate versus threshold for probability of detection}
To investigate the performance of the proposed scheme in achieving rate fairness, we plot the achievable minimum rate of the system in Fig. \ref{fig:R6_1}. Note that the MMR-max scheme acts as an optimal scheme for the achievable minimum rate due to its inherent power allocation based on maximizing the minimum CU rate.
As shown in the figure, the proposed GMR-max-based design yields a performance in close proximity to that of the optimal MMR-max scheme and improved over the SR-max scheme for both $N_\mathrm{RF}=\{16,~8\}$. This verifies the suitability of the proposed MM-Alt scheme for achieving rate fairness via the GMR-max framework in mmWave ISAC systems.

\subsubsection{Achievable sum rate versus threshold for probability of detection}
In Fig. \ref{fig:R7_1}, we plot the achievable sum rate of the system using the proposed algorithms. As shown in the figure, the SR-max scheme acts as an optimal scheme for the achievable sum rate of the system. This is due to the allocation of higher power to the stronger CUs possessing higher-quality channels. Interestingly, the performance of the GMR-max scheme is closer to that of the SR-max scheme and exceeds that of the MMR-max method for both $N_\mathrm{RF} = 8$ and $16$. This reveals that the MM-Alt algorithm proposed for the GMR-max scheme achieves rate fairness, without significantly compromising the achievable sum rate. Therefore, GM rate maximization is eminently suited for an optimal trade-off between the achievable sum rate and rate-fairness in the mmWave ISAC system.

It is worth noting that Fig. \ref{fig:R6_1} shows that the GMR-max outperforms the SR-max metric in fairness under the MMR-max framework, while Fig. \ref{fig:R7_1} demonstrates its superiority over the MMR-max in maintaining higher throughput under the SR-max framework. These results confirm that the GM rate metric ensures fairness without significantly degrading overall system performance, making it a well-balanced and effective metric for the mmWave ISAC systems.

\subsubsection{Achievable GM rate versus transmit power $P_\mathrm{t}$}


In Fig. \ref{fig:R8_1}, we plot the achievable GM rate of the system versus transmit power $P_\mathrm{t}$ for a fixed value of the probability of detection threshold, namely, $P = 0.975$ and for the probability of false alarm $P_\mathrm{FA} = 10^{-6}$. 
As expected, the achievable GM rate of the system increases with $P_\mathrm{t}$ due to having higher SINRs for the CUs. 
For comparison, we evaluate the proposed MM-Alt method against the WMMSE-Alt approach from \cite{mmWave_ISAC_1}. The WMMSE-Alt method reformulates the non-convex problem (\ref{GM_OP:4}) into a convex one using the WMMSE framework, which is then solved via alternating optimization. As illustrated in Fig. \ref{fig:R8_1}, the MM-Alt demonstrates superior performance by effectively handling non-convexity, resulting in improved efficiency in power distribution and overall system performance.

\subsubsection{Achievable GM rate versus probability of false alarm}
Fig. \ref{fig:R9_1} plots the achievable GM rate versus the probability of false alarm $P_\mathrm{FA}$ for the fixed probability of detection $P_\mathrm{th}=0.975$ and transmit power $P_\mathrm{t}=30$ dBm. As shown in the figure, the achievable GM rate of the system improves upon increasing $P_\mathrm{FA}$ from $10^{-6}$ to $10^{-1}$. This is due to the fact that the power radiated towards the RT decreases upon increasing $P_\mathrm{FA}$, which results in a higher SINR at the CUs. 
Additionally, observe that the MM-Alt consistently achieves a higher GM rate across different probabilities of false alarm, demonstrating its effectiveness in optimizing the sensing-communication trade-off.
Furthermore, the performance of the GMR-max scheme having $N_\mathrm{RF}=16$ achieves levels coinciding with its FDB counterpart for low as well as high values of $P_\mathrm{FA}$. This confirms the efficiency of the proposed MM-Alt scheme. Thus, the proposed MM-Alt algorithm is a power-efficient method conceived for beamforming optimization in ISAC-aided mmWave systems due to its requirement of fewer RFCs, while also achieving the GM rate of the FDB scheme.

\subsubsection{Achievable GM rate versus number of CUs}
Fig. \ref{fig:R15_111} shows the achievable GM rate decreasing as the number of CUs increases from $2$ to $8$, with a fixed transmit power of $30$ dBm and $8$ RFCs. This decline is due to CUs competing for limited resources, reducing per-CU beamforming gain. However, the proposed MM-Alt algorithm with fewer RFCs performs close to the FDB scheme even for a higher number of users, highlighting its scalability and efficiency in multi-user hybrid beamforming.

\section{\uppercase{Conclusion}}\label{conclusion}
In this paper, we conceived novel HBF designs for optimizing the BB and RF TPCs for joint communications and sensing by exploiting the spatial degrees of freedom in the mmWave ISAC system. To evaluate the sensing and communication performances, a pair of problems, namely: PD-max and GMR-max, were formulated considering the QoS of the RT and CUs, transmit power, and the unity magnitude constraints. A pair of power-efficient Bi-Alt and MM-Alt algorithms were proposed for solving the PD-max and GMR-max problems, respectively, which involve the SCA and PRCG algorithms for optimizing the RF and BB TPCs. 
Finally, simulation results were presented, which verify that our proposed design approaches the performance of the ideal yet impractical FDB, despite using a low number of RFCs. Furthermore, the proposed design outperformed various benchmark schemes, which shows the efficacy of the proposed algorithms. 
Moreover, extending the proposed design to multiple-RT scenarios presents additional challenges, requiring advanced HBF algorithms and refined sensing performance evaluations, which we leave for future work.

\bibliographystyle{IEEEtran}
\bibliography{biblio.bib}

\end{document}